\PassOptionsToPackage{table}{xcolor}
\documentclass[]{article} 
\usepackage{arxiv}

\usepackage[utf8]{inputenc} 
\usepackage[T1]{fontenc}    
\usepackage[english]{babel}
\usepackage{hyperref}       
\usepackage{url}            
\usepackage{doi}
\usepackage{microtype}      
\usepackage[toc,page]{appendix}

\usepackage{graphicx}
\usepackage{tikz}
\usepackage{tcolorbox}
\usepackage{pgfplots}
\usetikzlibrary{patterns, arrows, shapes}
\usetikzlibrary{intersections, pgfplots.fillbetween}
\pgfplotsset{compat=1.18} 

\usepackage{caption}
\usepackage{subfig}
\usepackage{rotating}

\usepackage{dsfont, mathrsfs, bm}
\usepackage{nicefrac}       
\usepackage{amsmath, amssymb, amsthm, mathtools}
\usepackage{threeparttable}
\newcolumntype{L}{>{\raggedright\arraybackslash}X}

\usepackage[ruled,linesnumbered]{algorithm2e}

\usepackage{makecell, multirow, tabularx}
\usepackage{tabularray}
\usepackage{booktabs}       

\newcommand{\rowgroup}[1]{\hspace{-0.4em}#1}

\usepackage{todonotes}
\usepackage{cleveref}
\Crefname{figure}{figures}{Figures}

\usepackage[round, authoryear]{natbib}
\usepackage{pifont}

\definecolor{rptublaugrau}{RGB}{80,114,137}
\definecolor{rptugruengrau}{RGB}{119,182,186}
\definecolor{rptudunkelblau}{RGB}{4,44,88}
\definecolor{rptuhellblau}{RGB}{106,178,231}
\definecolor{rptudunkelgruen}{RGB}{0,107,107}
\definecolor{rptuhellgruen}{RGB}{38,208,124}
\definecolor{rptuviolett}{RGB}{76,53,117}
\definecolor{rptupink}{RGB}{209,56,150}
\definecolor{rpturot}{RGB}{227,27,76}
\definecolor{rptuorange}{RGB}{255,162,82}
\definecolor{rptuschwarz}{RGB}{0,0,0}
\definecolor{rptuweiss}{RGB}{255,255,255}

\title{Goodness-of-fit tests for spatial point processes:\\A power study}

\date{May 19, 2025}	

\author{Chiara~Fend\\
	Department of Mathematics\\
	RPTU University Kaiserslautern-Landau\\
	Kaiserslautern, Germany \\
	\texttt{chiara.fend@rptu.de} \\
	\And Claudia~Redenbach \\
	Department of Mathematics\\
	RPTU University Kaiserslautern-Landau\\
	Kaiserslautern, Germany \\
	\texttt{claudia.redenbach@rptu.de}
}

\renewcommand{\shorttitle}{GOF Tests for spatial point processes: A power study}

\hypersetup{pdftitle={Goodness-of-fit tests for spatial point processes: A power study},
            pdfsubject={},
            pdfauthor={Chiara~Fend, Claudia~Redenbach},
            pdfkeywords={Point process, goodness-of-fit, tda}
}

\newcolumntype{C}{>{\centering\arraybackslash}X}

\DeclareMathOperator{\R}{\mathbb{R}}
\DeclareMathOperator{\N}{\mathbb{N}}

\newcommand{\1}[1]{\mathds{1}\!\left(#1\right)}
\newcommand{\abs}[1]{\left\lvert #1 \right\rvert}
\DeclarePairedDelimiterX{\norm}[1]{\lVert}{\rVert}{#1}

\DeclareMathOperator{\X}{\mathbf{X}}        
\DeclareMathOperator{\Y}{\mathbf{Y}}        
\DeclareMathOperator{\x}{\mathbf{x}}        

\newcommand{\cmark}{\textcolor{rptudunkelgruen}{\ding{51}}}%
\newcommand{\xmark}{\textcolor{rpturot}{\ding{55}}}%
\newcommand{\bestmark}{\textcolor{rptuorange}{\ding{72}}}%

\begin{document}
\maketitle


\begin{abstract}
	Spatial point processes are used as models in many different fields ranging from ecology and forestry to cosmology and materials science. In recent years, model validation, and in particular goodness-of-fit testing of a proposed point process model have seen many advances. Most of the proposed tests are based on a functional summary statistic of the observed pattern. In this paper, the empirical powers of many possible goodness-of-fit tests that can be constructed from such a summary statistic are compared in an extensive simulation study. Recently introduced functional summary statistics derived from topological data analysis and new constructions for the test statistic such as the continuous ranked probability score are included in the comparison. We discuss the performance of specific combinations of functional summary statistic and test statistic and their robustness with respect to other tuning parameters. Finally, tests using more than one individual functional summary statistic are also investigated. The results allow us to provide guidelines on how to choose powerful tests in a particular test stetting.
\end{abstract}



\section{Introduction}

In \citet{fend2025}, a general framework for the goodness-of-fit testing of spatial point processes was introduced. The reviewed goodness-of-fit tests are based on functional summary statistics of the point pattern. Several choices for these statistics are available, going from classical approaches such as Ripley's $K$-function to recent descriptors derived from topological data analysis. Each test in the framework is composed of a functional summary statistic, a test statistic, and an ordering that measures how extreme an observation is. 
Here, we present an extensive empirical investigation of the power of goodness-of-fit tests that fit the framework of \citet{fend2025}. In particular, we include combinations of the individual components that have, to the best of our knowledge, not yet been considered. We compare only Monte Carlo tests, i.e. we use simulations of the null model and their empirical functional summaries to decide if the observed value of the test statistic is extreme or not. Asymptotic theory for point processes has seen many developments in the last years, but it is still limited to restricted classes of point processes and particular summary functions and test statistics. This is the reason why we excluded asymptotic tests from this investigation. 

Several previous power studies on goodness-of-fit testing for spatial point processes were reviewed in Section~7 of \citet{fend2025}. Although most studies concentrate on the null hypothesis of complete spatial randomness (CSR), the general setting differs widely, as very different models and parameter choices were used to simulate the observed data. Consequently, combining and comparing results from these studies is not straightforward.
In addition, many possible tests that can be built from the general framework have to be tested. Examples are the use of the continuous ranked probability score as test statistic or topological data analysis based summary statistics in the Monte Carlo setting.

The main focus of our study is on the combination of summary statistic, test statistic, and ordering. We test the null hypothesis of complete spatial randomness, as this is the most common hypothesis in the literature. In order to investigate the empirical power of the test for different types of deviations from CSR, we consider five point process models, namely a Matérn cluster point process, the Baddeley-Silverman cell process, a Hardcore process, a Gaussian determinantal point process and a Strauss process.  
 
The reminder of this work is organized as follows. In Section~\ref{sec:gof} we
provide a small introduction to the individual components of the goodness-of-fit tests. Recommendations obtained from prior studies and remaining open questions are summarized in Section~\ref{sec:prior}. The set up of our simulation study is described in Section~\ref{sec:study}. Section~\ref{sec:result} presents the results of the simulation study. The paper concludes with a discussion of the open questions in Section~\ref{sec:discussion}.

\section{General framework for goodness-of-fit tests}\label{sec:gof}

In this section we briefly present the framework for goodness-of-fit tests that was introduced in \citet{fend2025}. For a detailed introduction we refer the reader to this review article. 


Let $\X$ be a spatial point process on $\R^d$, $d > 1$. We assume that $\X$ is simple, which means that the points are almost surely pairwise distinct. The point process is then given as the random set $\X = \{X_1, \dots, X_N\}$ where $N \in \N \cup \{\infty\}$. Additionally, we assume stationarity and isotropy of the point process, which means that the distribution of the process is invariant under translations and rotations around the origin. The intensity $\lambda > 0$ of a stationary point process $\X$ is the expected number of points per unit volume. We have $\mathbb{E}\left[\X(B)\right] = \mathbb{E}\left[\#\left(\{X_1, \dots, X_N\} \cap B\right)\right] = \lambda \abs{B}$ for every Borel set $B \in \mathcal{B}^d$, where $\abs{\cdot}$ denotes the Lebesgue measure on $(\R^d, \mathcal{B}^d)$ and $\#A$ the cardinality of the countable set $A$.

A realization of $\X$ is a so-called point pattern $\x$. We observe $\x$ only within a bounded observation window $W \subset \R^d$ with $0 < \abs{W} < \infty$. The observed pattern is denoted as $\x_0 = \x \cap W = \{x_1, \dots, x_n\} $ for $n \in \N$.

In the following, let $B_r(x) = \{y \in \R^d \mid \norm{y-x} \leq r\}$ be the Euclidean ball with radius $r$ centered in $x \in\R^d$. The indicator function of a set $A$ is denoted as $\1{\cdot \in A}$. 


Based on the observation $\x_0$ of the point process $\X$ with unknown distribution $P$ we test the hypothesis \begin{equation*}
	H_0: P = P_0 \quad \text{vs.} \quad H_1: P \neq P_0.
\end{equation*}
For simplicity, we assume that $P_0$ is a fully specified point process model, which means that no parameters have to be estimated from the observed pattern. We refer the reader to \citet{fend2025} for the case of composite hypotheses.

In general, a test will be constructed using the following components: \begin{enumerate}
    \item A functional summary statistic $T(\X, \cdot):\mathcal{R} \to \R$ where $\mathcal{R} \subset \R$ and its nonparametric estimator $\widehat{T}(\x_0, \cdot):\mathcal{R} \to \R$.
    \item A test statistic $D$ that aggregates $\widehat{T}(\x_0, \cdot)$ on the relevant subset $\mathcal{R}^*$ of the domain $\mathcal{R}$. The test statistic $D$ can be scalar or vector-valued and can potentially be based on pointwise deviations from the null model.
    \item An ordering $\preceq$ on the range of $D$ that defines what is considered extreme under the null hypothesis. In our notation $D_k \preceq D_l$ means that $D_k$ is at least as extreme as $D_l$, thus small values indicate extremeness.
    \item A method for computing the critical value, either via asymptotic reasoning or by Monte Carlo simulation.
\end{enumerate}

Not every test statistic is suitable for every ordering such that we introduce the possible choices always in pairs. As mentioned in the introduction, we restrict our study to Monte Carlo tests. 

For this testing procedure, we sample $m$ point patterns $\x_1, \dots, \x_m$ from $P_0$. Let $D_0, \dots, D_m$ denote the values of the selected test statistic for $\x_0, \dots, \x_m$. 
We reject $H_0$ if the Monte Carlo $p$-value defined as 
\begin{equation*}
	p_{\operatorname{MC}} = \frac{1}{m+1} \left(1 + \displaystyle\sum_{i=1}^m \1{D_i \preceq D_0} \right)
\end{equation*}
is smaller than or equal to the selected significance level $\alpha \in (0,1)$.

From the literature, it is well known that a single summary statistic cannot fully describe the distribution of a point process. The two-step procedure of global envelope tests within the R-package \texttt{GET} \citep{GET} allows to combine $1 < K < \infty$ different individual functional summary statistics into one overall goodness-of-fit test. 

In the first step, we proceed with components 1 - 3 from above for each individual summary statistic $T_1, \dots, T_K$. For every summary statistic $T_k$, we summarize the result of the ordering in a scalar number $d_k$. This can be done, e.g., in terms of the scalar-valued test statistic or via the value of the depth measure that induces the ordering of a vector-valued test statistic. Note that we require that for all individual values $d_1, \dots, d_K$ only small values are considered extreme. 

In the second step, we define the vector $D = [d_1, \dots, d_K]^T \in \R^K$ as final test statistic. Then we continue with components 3 and 4 of the Monte Carlo test with the vector-valued test statistic. The ordering $\preceq$ in the two-step procedure is always induced by the extreme rank length measure \citep{myllymaki2017} with one-sided pointwise extremeness. More details on the two-step procedure and alternative approaches to combining functional summary statistics in a goodness-of-fit test setting can be found in \citet[Section~6.4]{fend2025}.

\subsection{Functional summary statistic}

There exist several functional summary statistics that can be used for goodness-of-fit tests and that consider different properties of the point process. A detailed theoretical introduction into the classical (probabilistic) functional summary statistics can be found in \citet{spatstatBuch} and \citet{moller2003}. The topological functional summary statistics are introduced in \citet{fend2025}.

The list of the functional summary statistics that are of interest in our simulation study in Section~\ref{sec:study} as well as the nonparametric estimators that we used can be found in Table~\ref{tab:sum-funs}. To understand the construction of the estimators, further notation is necessary, which we introduce below. 

The domain $\mathcal{R}$ of all eleven functional summary statistics is given as a subset of $\R_{\geq 0}$. The evaluation point $r \in \mathcal{R}$ then denotes a spatial range or distance within the point pattern.

In case of Ripley's $K$-function and the $L$-function we use Ripley's isotropic edge correction estimator. 
The edge correction factor $p_W(x_1, \norm{x_1 - x_2})$ is the probability that given the first point $x_1$, the second point $x_2$ lying on $\partial B(x_1, \norm{x_1-x_2})$ falls into the observation window $W$. The exact formula of $p_W(x_1, \norm{x_1 - x_2})$ can be found in \citet[Sec.~7.4.4]{spatstatBuch}. For the transformation of the $K$-function into the $L$-function we additionally need the volume $\omega_d$ of the $d$-dimensional unit ball, e.g., if $d=2$ we have $\omega_2 = \abs{B_1(0)}= \pi$.

Since the pair-correlation function is a density, it is estimated using a kernel estimator. The function $k_b$ hereby denotes the Epanechnikov kernel with bandwidth $b$, see \citet[Sec.~7.6.2]{spatstatBuch}. Furthermore, the factor $\sigma_d$ denotes the surface area of the ball with radius $1$ in $\R^d$, e.g., $\sigma_2 = 2\pi$.

In case of the distance-based functional summary statistics $F$, $G$ and $J$, we use the Kaplan-Meier style edge corrections discussed in \citet[Sec.~8.11.4]{spatstatBuch}. For the nearest neighbor distance distribution function $G$ let $d_{nn}(x_i) = d(x_i, \x_0\setminus \{x_i\})$ be the distance from $x_i$ to its nearest neighbor in $\x_0$. The so-called observed failure distance of $x_i$ is defined as $t(x_i) = \min \{ d_{nn}(x_i), d(x_i, \partial W) \}$. 
For estimation of the empty space function, the points $x_i$ are replaced by test locations $u$ in a regular lattice $\operatorname{Lat} \subset W$. A test position $u \in \operatorname{Lat}$ takes on the role of the observed point $x_i$ in the $G$-function estimator. Thus, we have $d_{nn}(u) = d(u, \x_0)$ and $t(u) = \min \{ d_{nn}(u), d(u, \partial W) \}$.

For the five topological functional summary statistics, we use the alpha-complex filtration of the observed point pattern $\x_0$ which is a filtration of the Delaunay triangulation. The alpha-complex at distance $r > 0$ is given as 
\begin{equation*}
\mathcal{K}_r = \{ \sigma \text{ simplex with vertices in } \x_0 \mid \bigcap_{x_i \in V(\sigma)} (B_r(x_i) \cap \mathrm{Voronoi}(x_i, \x_0)) \; \neq \emptyset \} \end{equation*} 
where $\mathrm{Voronoi}(x_i, \x_0)$ denotes the Voronoi cell of the point $x_i$ with respect to the entire point pattern $\x_0$ and $V(\sigma)$ the set of vertices of the simplex. The set of $p$-dimensional topological features of the filtration is denoted by $J^p$ for $p \in \N_0$. The $0$-dimensional features are the connected components, the $1$-dimensional features are circular loops, and the $2$-dimensional features are voids in space. Each topological feature $j$ has a birth time $b_j$ that corresponds to the time it first appeared in the filtration and a death time $d_j$ corresponding to the last occurrence. 

From the functional summary statistics introduced in \citet{fend2025}, we do not include the variance-stabilizing $\arcsin$-transformation of $G$ called $G^{\star}$ as it is rarely used in practice. The topological summary function $\operatorname{ND}_0$ is also excluded as it is by definition the complement of the $0$-dimensional Betti curve, the latter being the better known standard topological characteristic.

\begin{table*}[th]\centering
\renewcommand{\arraystretch}{2.}
\rowcolors{3}{gray!10}{white}
\begin{tabularx}{\textwidth}{lLl@{\hskip 15pt}r} \toprule
\hiderowcolors 
$T$ & Name & Estimator $\widehat{T}(\x_0, r)$ & Equation \\ \midrule 
\showrowcolors
$K$ & Ripley's $K$-function & $\frac{\abs{W}}{n(n-1)}\displaystyle\sum_{x_1, x_2 \in \x_0}^{\neq} \frac{\1{\norm{x_1-x_2} \leq r}}{p_W(x_1, \norm{x_1 - x_2})}$& (1)\\
$L$ & $L$-function & $\left(\widehat{K}(\x_0, r)/\omega_d\right)^{\frac{1}{d}}$ & (2)\\
$pcf$ & pair correlation function & $\frac{\abs{W}}{\sigma_d r^{d-1}n(n-1)}\displaystyle\sum_{x_1, x_2 \in \x_0}^{\neq} \frac{k_b\!\left(r-\norm{x_1-x_2}\right)}{p_W(x_1, \norm{x_1-x_2})}$ & (4)\\
$F$ & Empty space function & $1 - \displaystyle\prod_{s \leq r} \left( 1-\frac{\#\{ u \in \operatorname{Lat} \mid t(u) = s, d_{nn}(u) \leq t(u_i) \}}{\#\{ u \in \operatorname{Lat} \mid t(u) \geq s \} }\right)$ & (5)\\
$G$ & Nearest neighbor distance distribution function &  $1 - \displaystyle\prod_{s \leq r} \left( 1-\frac{\#\{ x_i \in \x_0 \mid t(x_i) = s, d_{nn}(x_i) \leq t(x_i) \}}{\#\{ x_i \in \x_0 \mid t(x_i) \geq s \} }\right)$ &(6)\\
$J$ & $J$-function & $(1-\widehat{G}(\x_0, r))/(1-\widehat{F}(\x_0, r))$&(8)\\ \midrule
$\beta_0$ & $0$-dimensional Betti curve & $\displaystyle\sum_{j \in J^0} \1{b_j \leq r, d_j > r}$ &(14) \\
$\beta_1$ & $1$-dimensional Betti curve & $\displaystyle\sum_{j \in J^1} \1{b_j \leq r, d_j > r}$ &(14)\\
$\operatorname{APF}_0$ & $0$-dimensional accumulated persistence function & $\displaystyle\sum_{j\in J^0} d_j \1{d_j \leq r}$& (15) \\
$\operatorname{APF}_1$ & $1$-dimensional accumulated persistence function  & $\displaystyle\sum_{j\in J^1} (d_j-b_j) \1{b_j \leq r}$& (16) \\ 
$\chi$ & Euler characteristic curve & $\displaystyle\sum_{\sigma \in \mathcal{K}_r} (-1)^{\# V(\sigma)-1}$& (18)\\ 
\bottomrule 
\end{tabularx}
\caption{List of the functional summary statistics appearing in the simulation study. The equation numbers refer to the corresponding equation in \citet{fend2025} that introduces the theoretical counterpart.}
\label{tab:sum-funs}
\end{table*}

\subsection{Test statistic and ordering}

For the test statistic, we generally distinguish three categories in the framework. Type A statistics are scalar-valued and based on summarizing the absolute pointwise deviations between the empirical functional summary statistic of the observed point pattern and the reference value under the null hypothesis. The reference value is either the true functional summary statistic of the null model at that point or an estimate obtained from a set of simulations of the null model. Since type A test statistics involve absolute pointwise deviations to the null, only large values of the test statistic are extreme. In this category, we find the maximum absolute deviation \citep{diggle1979}, the integrated squared deviation \citep{diggle1979, cressie1993, loosmore2006}, the scaling approaches of \citet{myllymaki2015} and the integrated continuous ranked probability score \citep{heinrichmertsching2024}.

All test statistics that are scalar-valued and do not use pointwise deviations belong to type B. Extremeness is two-sided in this case, as both larger and smaller values than under the null are significant. A point evaluation of the summary function was used by \citet{ripley1977} and \citet{diggle1979} for pointwise envelopes. The integral of an empirical functional summary statistic is used in many asymptotic tests.

The third type C involves test statistics that are not scalar-valued. In most cases, the test statistic is vector-valued and contains the empirical functional summary statistic at a finite number of evaluation points. Extremeness of the vectors in a Monte Carlo test is measured using a statistical depth function that induces an ordering on the set of vectors that are obtained through the simulations. This category allows us to include the global envelope tests of \citet{myllymaki2017} in our framework. The second test statistic in this group is the pointwise continuous ranked probability score based on \citet{heinrichmertsching2024}.

Table~\ref{tab:test-statistics} provides a list of test statistics and the possible orderings that we used in our simulation study. The relevant part of the domain $\mathcal{R}^* \subset \mathcal{R} \subseteq \R_{\geq 0}$ over which one integrates or takes the supremum is in most cases an interval $[r_{\min}, r_{\max}]$. The \emph{optimal} choices for the bounds depend on the selected functional summary statistic, the type of observed point pattern, and the observation window. For several summary function estimators listed in Table~\ref{tab:sum-funs} there are computational restrictions due to denominators that could potentially be zero if the upper bound $r_{\max}$ is chosen too large. The estimator for the pair correlation function generally has a poor performance for $r$ close to $0$ \citep{spatstatBuch}. In that case, setting $r_{\min} > 0$ is necessary. For all other functional summary statistics, we can set $r_{\min} = 0$.
The only test statistic that does not require an interval $\mathcal{R}^*$ is the POINT test statistic, as it uses only a single evaluation point $r^*$.

\begin{table*}[th]\centering
\renewcommand{\arraystretch}{2.5}
\rowcolors{3}{gray!10}{white}
\begin{tabularx}{\textwidth}{lllL@{\hskip 15pt}r} \toprule
\hiderowcolors 
Type & Statistic & Formula & Ordering  & Equations \\ \midrule 
\showrowcolors
\cellcolor{white}& MAD & $\displaystyle\sup_{r \, \in \mathcal{R}^*} |\widehat{T}(\x_0, r) - T(\Y, r)|$ & larger & (19), (32) \\
\cellcolor{white}&DCLF & $\displaystyle\int_{\mathcal{R}^*} (\widehat{T}(\x_0, r) - T(\Y, r))^2 \; \mathrm{d}r$& larger & (20), (32)\\
\cellcolor{white}&ST & $\displaystyle\sup_{r \, \in \mathcal{R}^*} \left|\frac{\widehat{T}(\x_0, r) - T(\Y, r)}{\operatorname{sd}(\widehat{T}(\Y, r))} \right|$&larger & (21), (32) \\
\cellcolor{white}&QDIR & $ \!\begin{aligned}[t]
&\sup_{r \, \in \mathcal{R}^*} \Bigg\{ \mathds{1}\!\left(\widehat{T}(\x_0, r) \geq T(\Y, r)\right)\cdot \frac{\widehat{T}(\x_0, r) - T(\Y, r)}{|\widehat{T}(\Y, r)_{0.975} - T(\Y, r)|} \\ & \qquad\quad + \mathds{1}\!\left(\widehat{T}(\x_0, r) < T(\Y, r)\right)\cdot \frac{T(\Y, r)-\widehat{T}(\x_0, r)}{|\widehat{T}(\Y, r)_{0.025}- T(\Y, r)|}\; \Bigg\} \end{aligned}$&larger & (22), (32) \\
\multirow{-5}{*}{\cellcolor{white} A} & CRPS & $\!\begin{aligned}[t]
    &\displaystyle\mathbb{E}\!\left[\int_{\mathcal{R}^*} |\widehat{T}(\x_0, r) - \widehat{T}(\Y, r)| \; \mathrm{d}r \right] \\
    &\quad - \quad \frac{1}{2} \cdot\mathbb{E}\!\left[ \int_{\mathcal{R}^*} |\widehat{T}(\Y, r) - \widehat{T}(\Y^\prime, r)| \; \mathrm{d}r \right]
    \end{aligned}$&larger & (25), (32)\\
\midrule
\cellcolor{white}&INT & $\int_{\mathcal{R}^*} \widehat{T}(\x_0, r) \; \mathrm{d}r$ & two-sided &  (27), (32) \\
\multirow{-2}{*}{\cellcolor{white} B} & POINT & $\widehat{T}(\x_0, r^*)$  & two-sided   & (26), (32) \\
\midrule
\cellcolor{white}& FUN & $\widehat{T}(\x_0, \cdot)\big|_{\mathcal{R}^*}$ & erl  & (28), (38) \\
\cellcolor{white}& FUN & $\widehat{T}(\x_0, \cdot)\big|_{\mathcal{R}^*}$& area  & (28), (40) \\ 
\multirow{-3}{*}{\cellcolor{white} C} & FUN & $\widehat{T}(\x_0, \cdot)\big|_{\mathcal{R}^*}$ & cont & (28), (37) \\ 
\bottomrule 
\end{tabularx}
\caption{Overview of the test statistics and the corresponding orderings used in the simulation study. Equations refers to the corresponding equations in \citet{fend2025} that provide the formulas for both the test statistic and the ordering. The point processes $\Y$ and $\Y^\prime$ are independent and fulfill the null hypothesis, i.e. their distribution is $P_0$. The quantity $\widehat{T}(\Y, r)_{\beta}$ appearing in QDIR denotes the $\beta$-quantile of the distribution of $\widehat{T}(\Y, r)$.}
\label{tab:test-statistics}
\end{table*}

From the list of test statistics in \citet{fend2025}, we excluded the pointwise continuous ranked probability score test statistic SCORE, as it is computationally expensive to compute both in time and in memory. In addition, some preliminary experiments showed that it did not perform better than the FUN test statistic. We also excluded the two scaled DCLF test statistics QDIR,DCLF and ST,DCLF, as prior studies recommend using the scaled MAD variants \citep{myllymaki2015}. 

\section{Recommendation based on prior studies}\label{sec:prior}

In the literature, several power studies for goodness-of-fit tests for spatial point processes have been conducted. The common null hypothesis is complete spatial randomness. The setting of the individual studies is very diverse, but many combinations of functional summary statistic and test statistic have not yet been investigated. Nevertheless, many interesting conclusions can already be drawn. A detailed list of all key findings from these previous power studies is given in Section~7 of \citet{fend2025}. The main results are summarized in the following. 

We start with the results obtained using classical functional summary statistics.
For the $L$-function, \citet{baddeley2014} found that the DCLF test statistics led in most cases to more powerful tests compared to the MAD statistic. By construction of the Baddeley-Silverman cell process, its theoretical $K$- or $L$-function coincides with the one from the Poisson process and thus CSR. Tests constructed with either the MAD, DCLF or INT test statistic are not able to distinguish the Baddeley-Silverman cell process from CSR. On the other hand, \citet{heinrichmertsching2024} were able to detect the difference when using the CRPS statistics even when it was combined with the $K$-function. 

The experiments in \citet{baddeley2014} showed that the DCLF and, to some extent, the MAD test statistics are sensitive to the choice of $\mathcal{R}^*$. \citet{myllymaki2017} found that FUN with the erl ordering, as well as QDIR and ST are more robust in terms of power when the upper bound $r_{\max}$ of $\mathcal{R}^*$ is varied.

The detailed discussion of test statistics of type A in \citet{myllymaki2015} states that using scaling approaches like QDIR and ST or variance-stabilizing transformations of the functional summary statistic not only makes the tests more robust when varying $r_{\max}$ but also increases their power compared to the raw version MAD. The authors conjecture that scaling alone should be sufficient in many settings.

Topological summary functions are known to yield powerful tests for the null hypothesis of CSR if the true model has complex interactions, e.g., in a Baddeley-Silverman cell process \citep{robins2016, biscio2019, biscio2020, botnan2022}. Nevertheless, \citet{krebs2022} discourage the use of a single topological summary statistics to draw conclusions, as the power greatly depends on the specific test setting. In particular, the use of $0$-dimensional or $1$-dimensional topological features can lead to very different decisions. An experiment in \citet{biscio2020} investigated the combination of two topological summary statistics which increased the power. Combined tests of the classical functional summary statistics were also already considered in \citet{mrkvicka2017}. Tests that combine $L$-, $F$-, $G$- and $J$-function were comparable in power to the most powerful of the individual tests.

The following concrete recommendations for the classical functional summary statistics are given in the literature:

\begin{itemize}
    \item If the range of interaction between points in the pattern is approximately known, then one should prefer DCLF over MAD with an upper bound $r_{\max}$ that is slightly larger than this range \citep{baddeley2014,spatstatBuch}. 
    \item QDIR can be recommended over ST and MAD in many settings when using classical functional summary statistics \citep{myllymaki2015, myllymaki2017}.
    \item The scaled MAD test statistics are often more powerful than the scaled DCLF test statistics \citep{myllymaki2015}.
    \item If extremeness is expected on a large part of $\mathcal{R}^*$, then one should prefer FUN with either the extreme rank length measure or the area measure. If extremeness is only expected at selected distances, then QDIR or FUN with the continuous rank measure are preferable \citep{mrkvicka2022, myllymaki2020, GET}.
\end{itemize}

\subsection{Open questions}\label{sec:open-question}

\paragraph{Performance of topological summary statistics}
The recommendations above are derived from experiments with the classical functional summary statistics. In case of the topological summary statistics, it is not known whether the same recommendations can be made. In contrast to the classical summary statistics, some of the topological summary statistics are by definition integer-valued instead of real-valued which can make a difference for the tests.

Additionally, to the best of our knowledge, a proper comparison between classical and topological functional summary statistics in a common goodness-of-fit test setting has not yet been published.

\paragraph{Necessary number of simulations in the Monte Carlo test}
An important aspect in Monte Carlo testing is the number of simulations $m$ of the null model. A discussion of this parameter can be found in \citet[Sec.~6.5]{fend2025}. In the literature, there are conflicting recommendations regarding the choice of $m$, ranging from $m=99$ to $m=2499$, depending on the choice of the test statistic. For some test statistics and orderings, the influence of $m$ has not been investigated in a point process setting, and no guidelines are available.

\paragraph{Use of the CRPS test statistic}
When using the $K$-function, the CRPS test statistic improved the power compared to, e.g., DCLF and MAD \citep{heinrichmertsching2024}. It remains open for which other summary statistics we can get a similar improvement.

\paragraph{Combination of topological and classical functional summary statistics}
Using individual topological functional summary statistics has been discouraged, while their combinations are mentioned to be worthwhile to investigate \citep{biscio2020}. Combinations of the classical functional summary statistics led to powerful tests in \citet{mrkvicka2017}. Combining both types of summary statistics in one test has not yet been considered and remains to be investigated. The knowledge about which combinations yield powerful tests then also provides more insight into in which way the topological characteristics can complement the classical probabilistic summary statistics.

\section{Experimental design}\label{sec:study}

We conducted several simulation studies in the case $d=2$, i.e. planar point processes, that provide answers to the open questions outlined above.

Assume that we observe the point pattern $\x_0$ that has $n$ distinct points in an observation window of the form $W_A = [0,\sqrt{A}]^2$ for some $A>0$.
Let $P$ denote the true distribution of the point process model producing the realization $\x_0$. The hypotheses are
\begin{equation*}
	H_0: P \text{ is complete spatial randomness} \quad \text{vs.} \quad H_1: P \text{ is not complete spatial randomness.}
\end{equation*}
Complete spatial randomness is modeled by the homogeneous Poisson process on $\R^2$. This process is parametrized by the intensity $\lambda > 0$, making the hypothesis a composite hypothesis. Thus, we proceed as in \citet[Sec.~6.1]{fend2025} and condition on the number of observed points in $W$. The null model $P_0$ from which we sample point patterns in the Monte Carlo test is then a binomial point process with a uniform intensity function on $W$. This means that we draw the $x$- and the $y$-coordinates of all $n$ points independently and uniformly in $[0,\sqrt{A}]$.

\subsection{Point process model families}\label{sec:models}

In this section, the point process models used in our study are introduced. All six processes are stationary and isotropic point processes. They cover different types of processes from clustering to repulsion between points. We will introduce the models only briefly. Details can be found in \cite{spatstatBuch}.

\paragraph{Homogeneous Poisson point process}
A point process $\X$ on $\R^2$ is a homogeneous Poisson process with intensity $\lambda > 0$ if the number of points $\X(B)$ in a Borel set $B \in \mathcal{B}^2$ is Poisson distributed with rate $\lambda \abs{B}$ and if for an arbitrary number $k$ of disjoint Borel sets $B_1, \dots, B_k$ we have that $\X(B_1), \dots, \X(B_k)$ are independent random variables. 

\paragraph{Matérn cluster point process}
The Matérn cluster point process is an example of a process that is used to model independent clustering in circular clusters of fixed radius $R > 0$. The parents, i.e., the center points of the clusters, are given by a homogeneous Poisson process with intensity $\kappa > 0$. For each parent point, we independently draw a number of children, which is Poisson distributed with rate $\mu > 0$. We place the children independently and uniformly in the disk with radius $R$ around the parent. Finally, the Matérn cluster point process is given as the union of the children of all parents. 

\paragraph{Baddeley-Silverman cell process}
The Baddeley-Silverman cell process is constructed by dividing the space into equal area cells. For each cell the random number $N_c$ of points within this cell is  drawn independently from the discrete distribution with probability weights 
\begin{equation*}
    P\!\left(N_c = 0\right) = \frac{1}{10}, \quad P\!\left(N_c = 1\right) = \frac{8}{9} \quad \text{and} \quad P\!\left(N_c = 10\right) = \frac{1}{90}.
\end{equation*}
Since $\mathbb{E}\!\left[N_c\right] = \operatorname{Var}(N_c) = 1$, we use $\lfloor\lambda\abs{W}\rfloor$ squares of equal size to obtain a process with intensity $\lambda>0$ in an observation window $W$ with equal side lengths. The points per cell are sampled independently and uniformly within the cell.

\paragraph{Hard core point process}
The hard core process with intensity parameter $\beta > 0$ and hard core radius $R > 0$ is a point process model in which two points are not allowed to be closer than distance $R$. This is realized using the density function 
\begin{equation*}
    f_{\text{hardcore}}(\x) \propto  \begin{cases} 
    \beta^{\#\!\x} & \text{if } \displaystyle\min_{x_1\neq x_2 \in \x} \norm{x_1-x_2} \geq R, \\
     0& \text{else}
    \end{cases}
\end{equation*}
w.r.t. the homogeneous unit intensity  Poisson process. This process can be seen as CSR under the restriction that the hard core distance is fulfilled.

\paragraph{Strauss point process}
The Strauss point process is a pairwise interaction process that is used to model repulsion between points. It is defined via the density function $f_{\text{strauss}}$ given as \begin{equation*}
    f_{\text{strauss}}(\x) \propto \beta^{\#\!\x} \prod_{x_1\neq x_2 \in \x} \gamma^{\1{\norm{x_1 - x_2} \leq R}}
\end{equation*}
where $\beta > 0$ is a parameter that controls the intensity, $\gamma \in [0,1]$ gives the strength of the interaction and $R > 0$ gives the range of interaction. 
For $\gamma=1$ we recover the homogeneous Poisson process and for $\gamma=0$ and the convention $0^0 = 1$ we obtain the hard core process.

\paragraph{Gaussian determinantal point process}
Determinantal point processes (DPP) form a class of point processes which are defined using a kernel $C$. If the $n$th order product density $\rho^{(n)}$ of spatial point process $\X$ on $\R^2$ is for all $n\in \N$ given as\begin{equation*}
    \rho^{(n)}(z_1, \dots, z_n) = \det \left(C(z_i, z_j)\right)_{1 \leq i,j \leq n} \quad \text{where} \quad z_1, \dots, z_n \in \R^2,
\end{equation*}
then we call $\X$ a (planar) DPP with kernel $C$. The Gaussian DPP uses the Gaussian kernel \begin{equation*}
    C(x_1, x_2) = \lambda \cdot \exp\left(-\norm{\nicefrac{(x_1 - x_2)}{\alpha}}^2\right)
\end{equation*} where $\lambda > 0$ is the intensity and the shape parameter $\alpha$ needs to fulfill $0 < \alpha < (\lambda\pi)^{-\frac{1}{2}}$ for the existence of the planar DPP \citep{lavancier_determinantal_2015}.

\subsection{General study parameters}

\begin{figure}
\centering
  	\includegraphics[height=0.9\textheight]{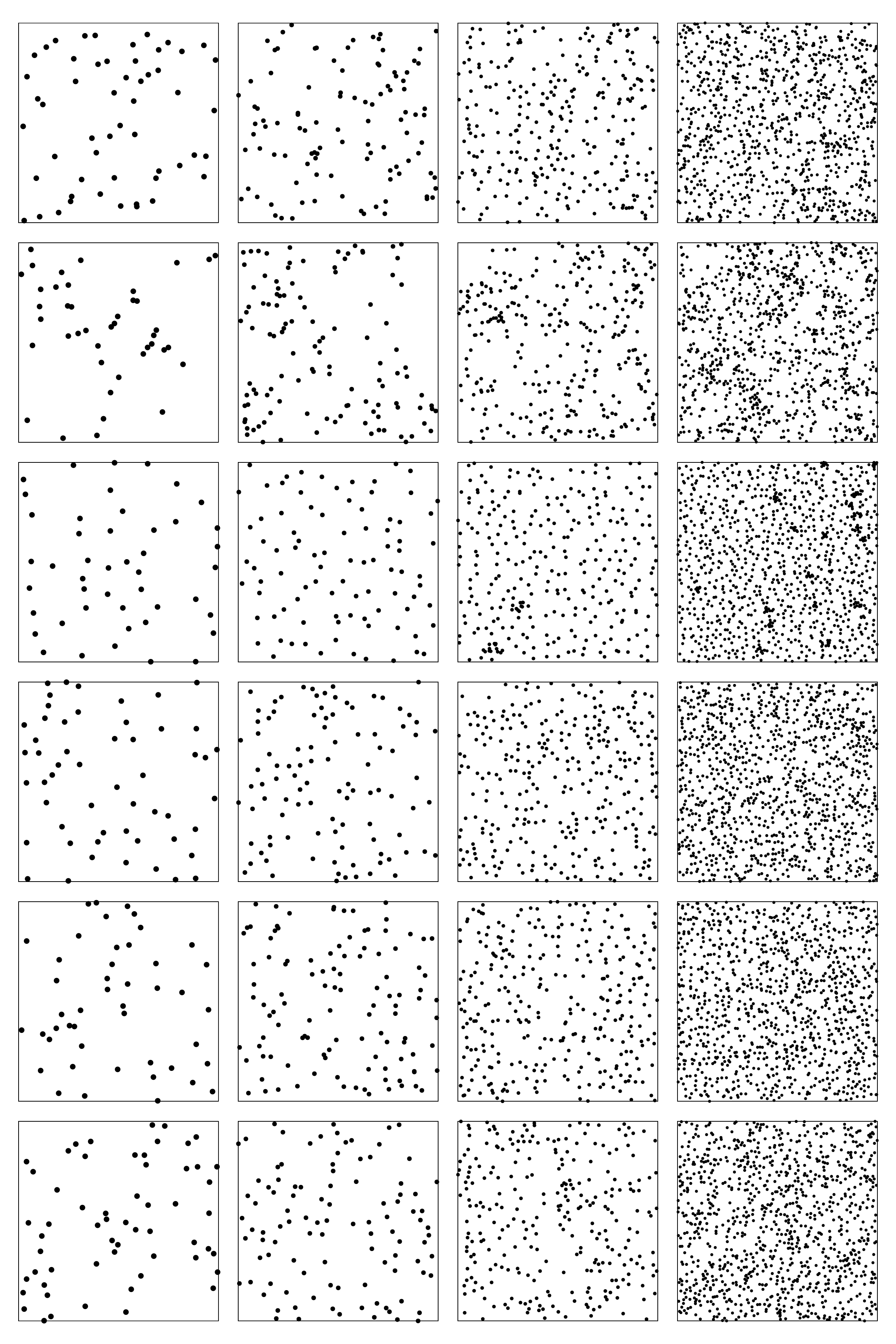}
	\caption{One realization of each of the point process models in Table~\ref{tab:parameter-setting} in each of the observation windows $W_A$. In the rows from top to bottom: Poi, MatClu, BadSil, Hard, Str, GDPP. In the columns from left to right: $W_1$, $W_2$, $W_6$, $W_{20}$.}
	\label{fig:models-examples}
\end{figure}

Table~\ref{tab:parameter-setting} states the exact parameters that were used in this study. The parameters were chosen such that the realizations visually resemble CSR, e.g., sparse clusters for the Matérn cluster point process and a shape parameter $\alpha$ which is close to the largest possible one for the Gaussian DPP. We simulated each of the six point processes $1000$ times on four different window sizes, namely $W_A = [0, \sqrt{A}]^2$ for $A \in \{1, 2, 6, 20\}$. Due to the intensity being set to $\lambda=50$ we obtain patterns with approximately $50$, $100$, $300$ and $1000$ points, respectively.
We used the R-package \texttt{spatstat} \citep{spatstat} for the simulation of all processes. One realization per model and observation window is shown in Figure~\ref{fig:models-examples}.

\begin{table*}[th]\centering
\renewcommand{\arraystretch}{1.5}
\rowcolors{3}{gray!10}{white}
\begin{tabularx}{\textwidth}{lLll} \toprule
\hiderowcolors 
Name & Point process family & Type &Parameters \\ \midrule
\showrowcolors
Poi & homogeneous Poisson point process & complete spatial randomness & $\lambda = 50$ \\
MatClu & Matérn cluster point process & clustered &$\kappa=25$, $\mu=2$, $R=0.2$ \\
BadSil & Baddeley-Silverman cell process & repulsive with few clusters & $\lambda=50$ \\
Hard & Hard core point process & hard core  &$\beta=80$, $R=0.05$ \\
Str & Strauss point process & repulsive &$\beta=95$, $\gamma=0.6$, $R=1$ \\ 
GDPP & Gaussian determinantal point process & repulsive & $\lambda=50$, $\alpha=0.05$ \\ \bottomrule
\end{tabularx}
\caption{Overview of the parameters used in the simulation study}
\label{tab:parameter-setting}
\end{table*}

We ran prior tests to see up to which maximal spatial distance $r_{\max} >0$ the estimation of each functional summary statistic was possible for all six specified point process models on all four window sizes. This resulted in a maximal upper bound of $r_{\max} = 0.1$ for the distance-based summary statistics $F, G$ and $J$ and $r_{\max}=0.25$ for the second-order and topological summary statistics. For all summary functions except the $pcf$, the relevant part of the domain $\mathcal{R}^*$ the interval $\mathcal{R}^* = [0, r]$ for some $0 < r \leq r_{\max}$. For the $pcf$, we use $\mathcal{R}^* = [0.005, r]$ with some $0 < r \leq r_{\max}$ as the estimation of the $pcf$ is unstable close to $0$. 
For a comparison involving the point evaluation test statistic POINT, we let $r^*$ coincide with the upper bound of $\mathcal{R}^*$.

We follow the default setting in \texttt{spatstat} for the estimation of the functional summary statistics and choose $513$ equidistant evaluation points in each $\mathcal{R}^*$. The empirical topological summary statistics are implemented using the R-package \texttt{TDA} \citep{TDA} while we used the \texttt{spatstat} implementation of the estimators of the classical functional summary statistics.

For the DCLF and MAD tests, we used the respective \texttt{spatstat} functions while QDIR, ST and FUN with all three types of orderings are available in the R-package \texttt{GET} \citep{GET}. Details regarding the implementation of the CRPS test statistic are given in \citet{fend2025}. The INT test statistic is approximated using a Riemann sum.

In the Monte Carlo tests we used $m \in \{99, 299\}$ for test statistics of type A and B and $m \in \{99, 299, 499, 999\}$ for type C. 

\section{Results} \label{sec:result}

In the following, we present a discussion of a selection of the empirical rejection rates computed for a fixed significance level of $\alpha=0.05$. Due to the high number of tests that were conducted and thus can be compared, interactive graphics are necessary to keep the amount of information manageable. These graphics allow the reader to selected the tests of interest. The interactive graphics are available in our repository \url{https://github.com/cfend/SpatialPointProcessesGOF}. This repository also contains the power plots for all four windows sizes.

\subsection{Individual functional summary statistics}\label{sec:result-indivi}

\begin{figure}
    \centering
     \subfloat{\includegraphics[width=\linewidth]{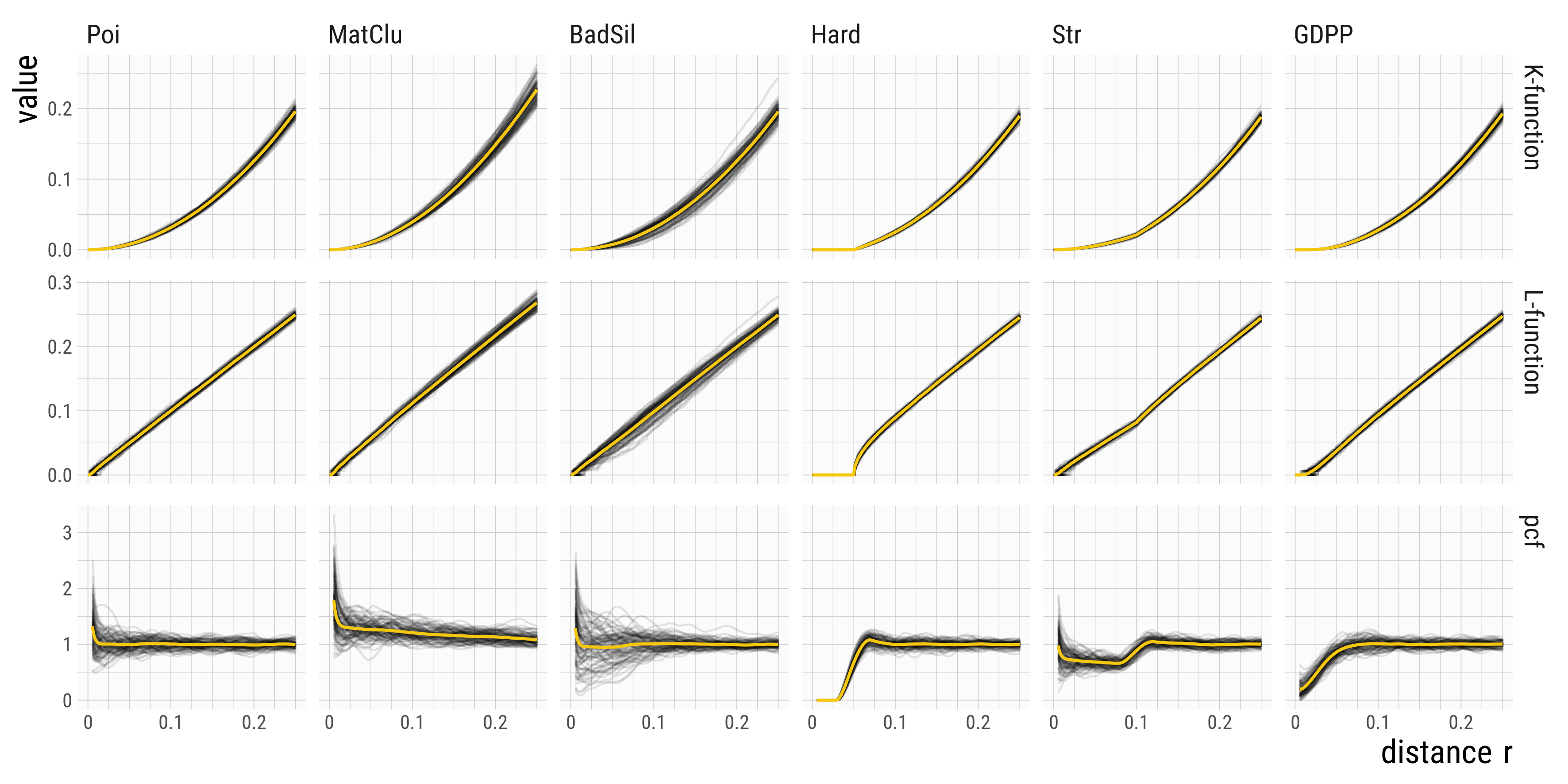}}
    \vspace{-3ex}
     \subfloat{\includegraphics[width=\linewidth]{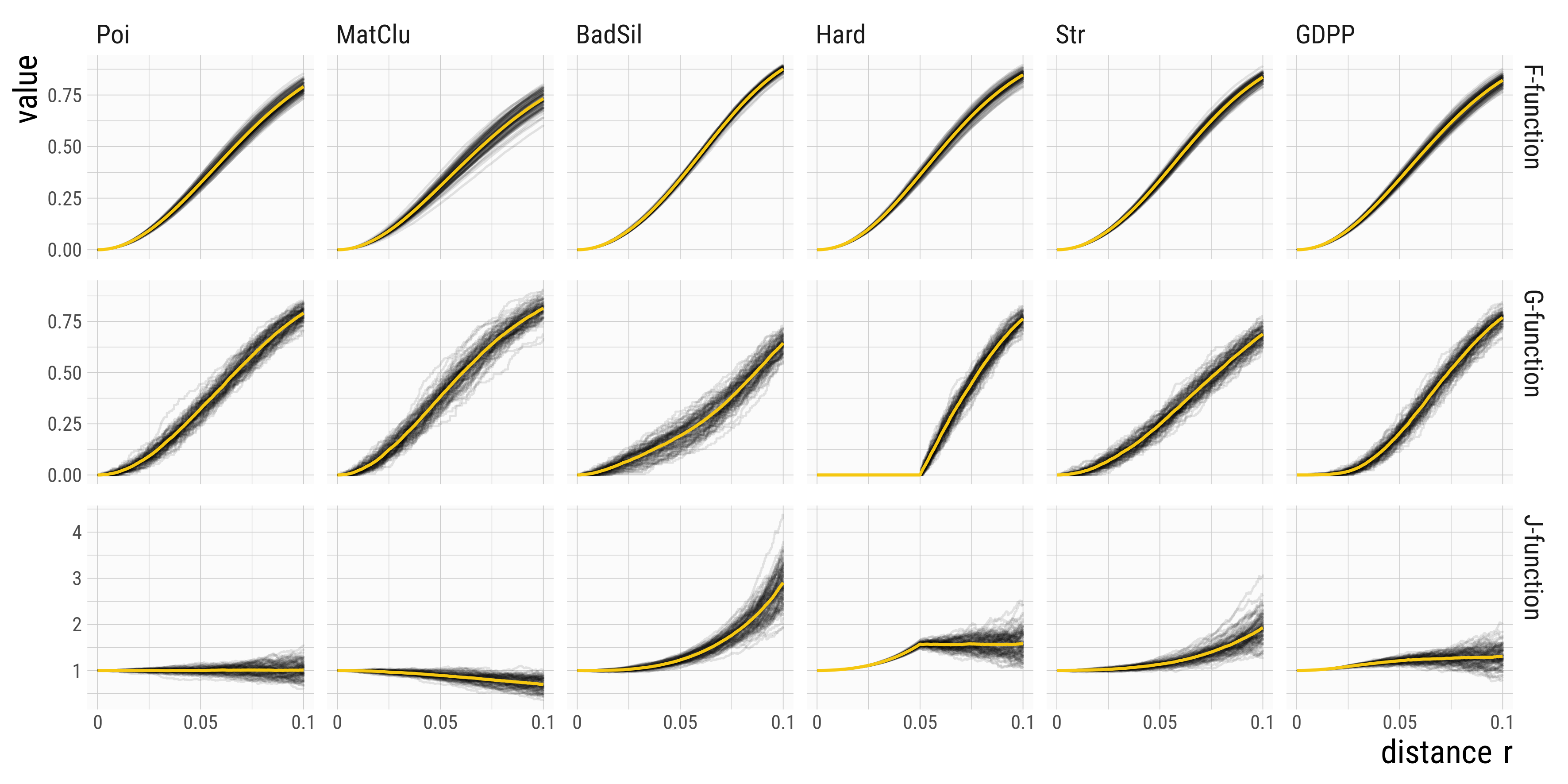}}
     
    \caption{Empirical classical functional summary statistics from $100$ realizations per model on window $W_6$. In yellow the pointwise mean of the individual estimates.}
    \label{fig:classical-means}
\end{figure}

\begin{figure}[th]
    \centering
    \includegraphics[width=\linewidth]{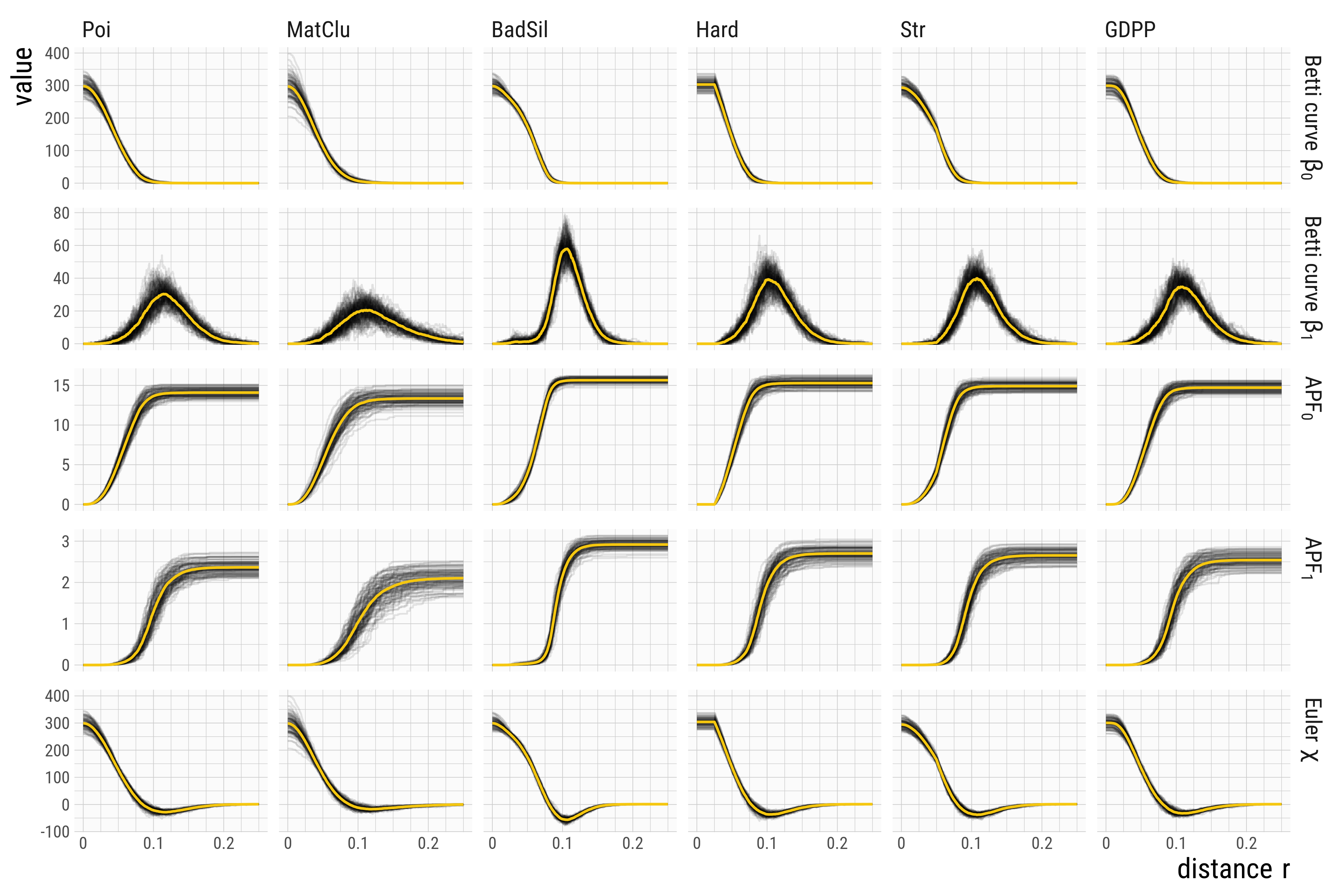}
    \vspace{0.5ex}
    
    \caption{Empirical topological functional summary statistics from $100$ realizations per model on window $W_6$. In yellow the pointwise mean of the individual estimates.}
    \label{fig:tda-means}
\end{figure}

In the first part of the study, we investigated the combination of a functional summary statistic and a test statistic. In the following, we show the results obtained for patterns in the observation window $W_{6}$ thus with approximately $300$ points per pattern. The number of simulations is fixed to $m=299$ unless mentioned otherwise. Examples of the empirical functional summary statistics for each model are shown in \Cref{fig:classical-means,fig:tda-means}. 

\subsubsection*{Test statistics DCLF, MAD, INT and POINT}

\begin{figure}[th]
    \centering
    \subfloat{\includegraphics[width=0.9\linewidth]{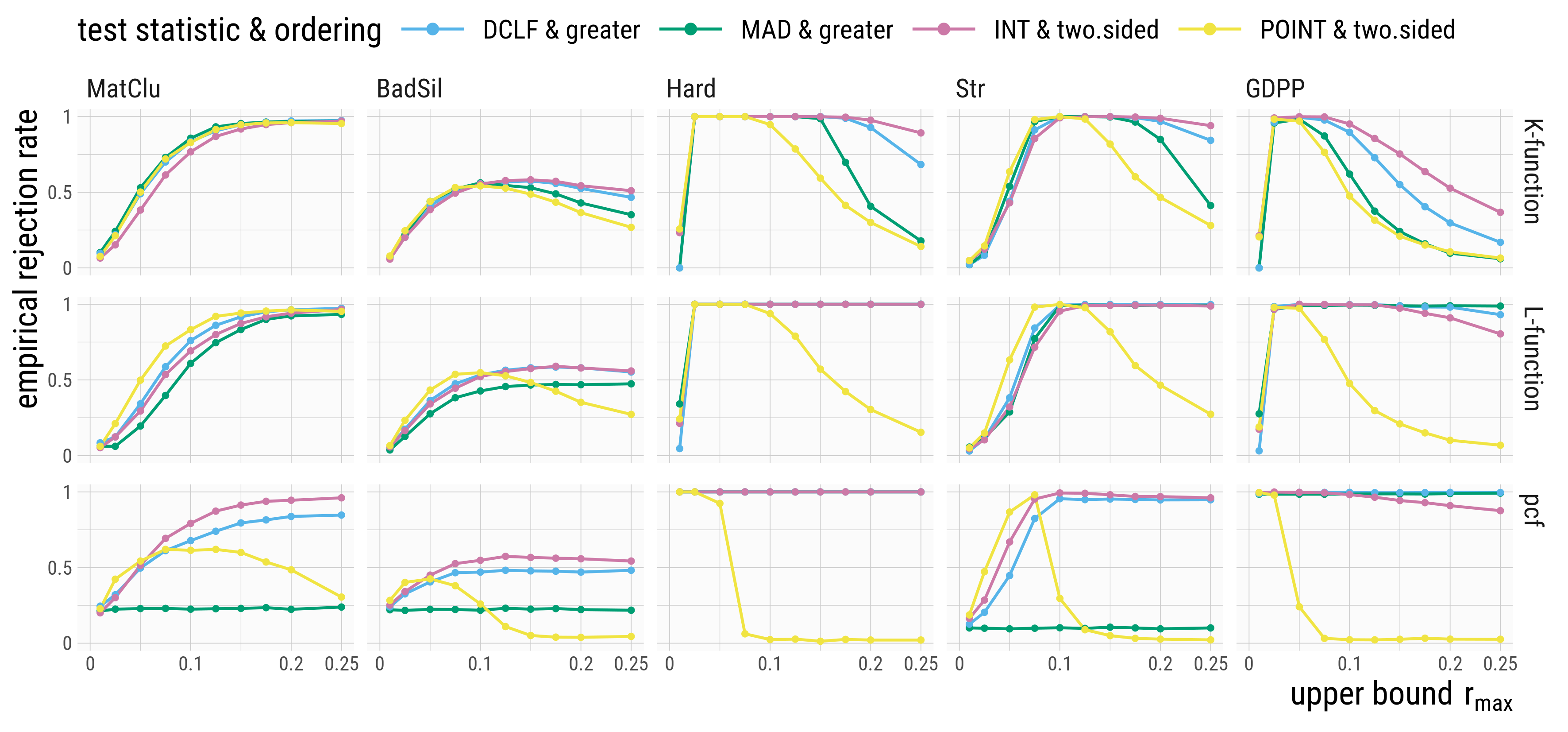}}
    \vspace{-4ex}
    \subfloat{\includegraphics[width=0.9\linewidth]{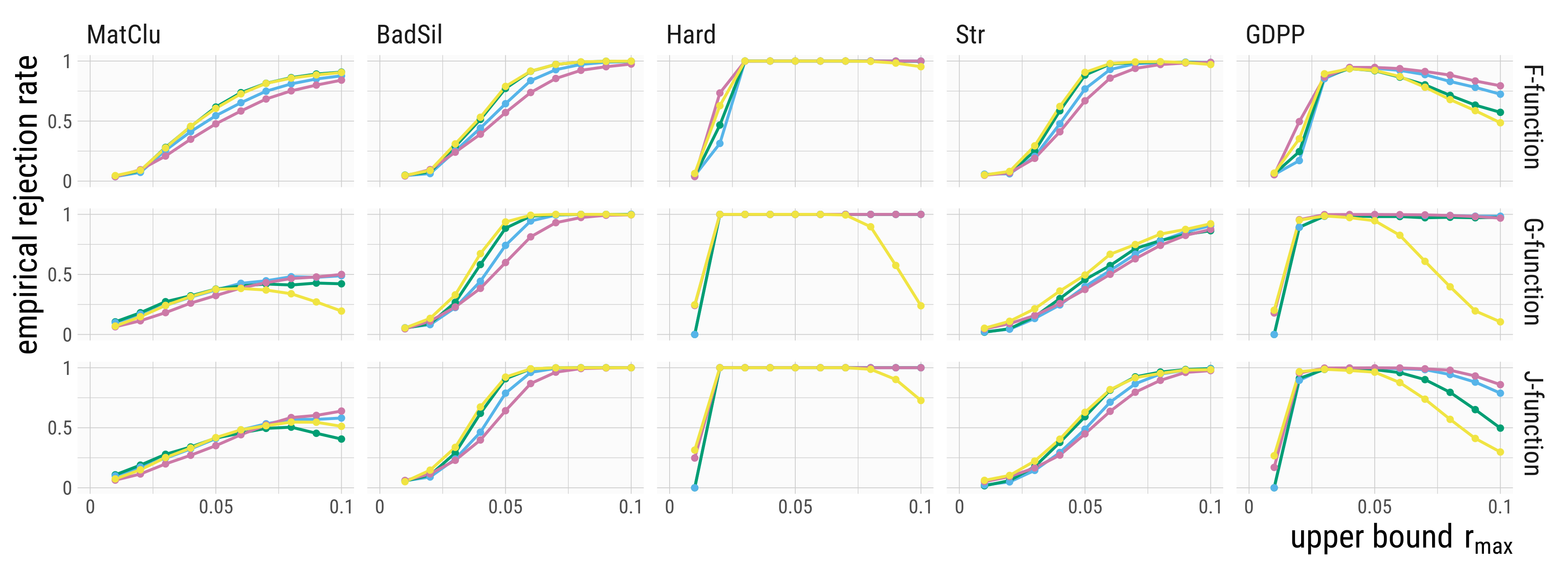}}
    \vspace{-4ex}
    \subfloat{\includegraphics[width=0.9\linewidth]{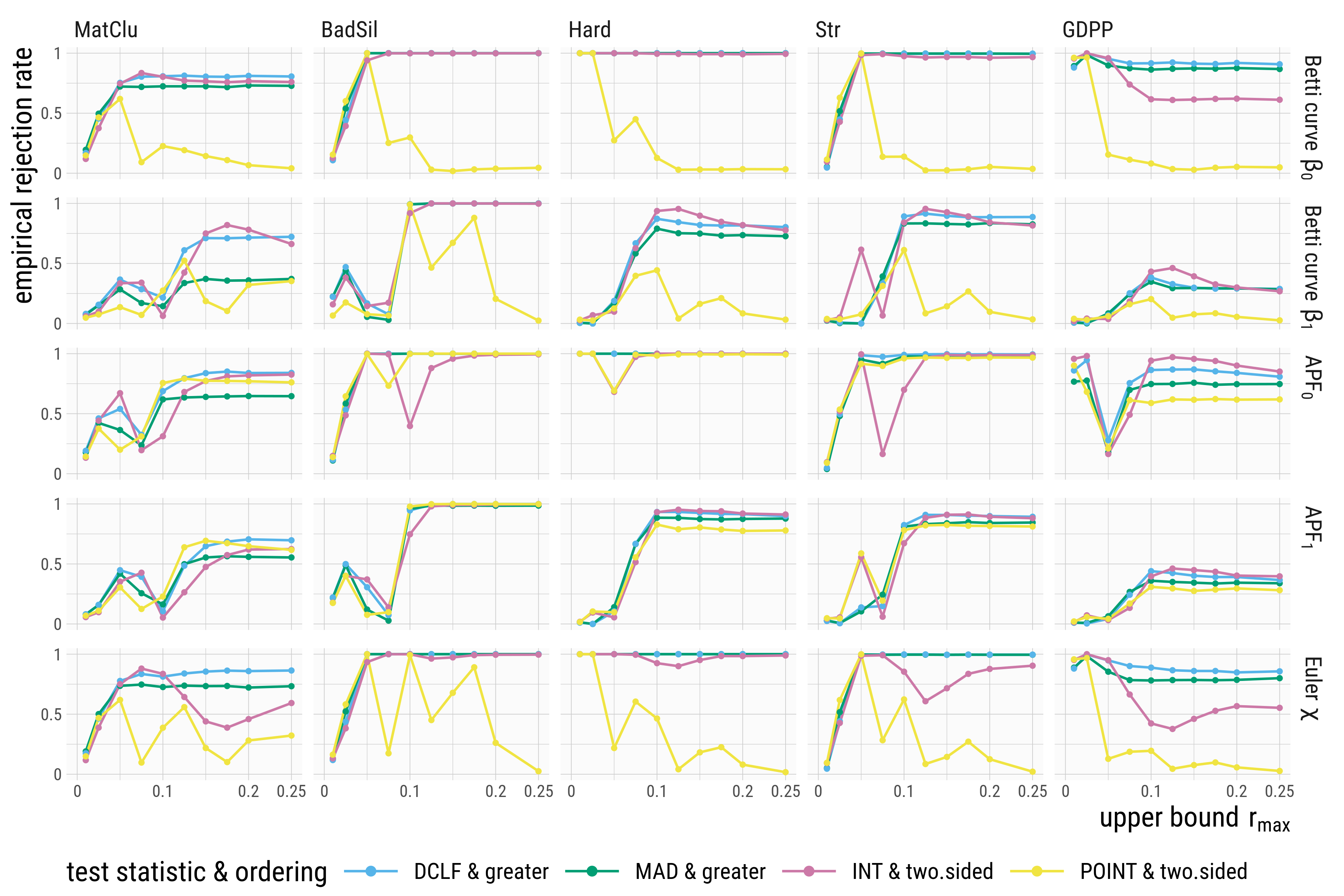}}

    \caption{Empirical power curves for the test statistics DCLF, MAD, INT and POINT with respect to the upper bound $r_{\max}$ of the domain $\mathcal{R}^*$. We used $m=299$ simulations on the observation window $W_6$.}
    \label{fig:dclf-mad-int}
\end{figure}

Figure~\ref{fig:dclf-mad-int} shows the empirical power curves for DCLF, MAD, INT and POINT in combination with the classical and topological functional summary statistics. The empirical sizes, which are not shown in the figure, met in all settings the significance level $\alpha=0.05$.

We first discuss the results obtained for the classical functional summary statistics. For the second-order summary statistics (Figure~\ref{fig:dclf-mad-int} first three rows from the top) we conclude that DCLF should be preferred over MAD. The only exception is the $L$-function when the alternative is the GDPP, which is a repulsive point process. This agrees with the experiments in \citet{baddeley2014}. 
The pure integral test statistic INT often results in higher empirical powers, especially for the pcf and the $K$-function. But, as expected, due to the high variability in the estimation of the $K$-function the tests are less robust with respect to the upper bound. The $L$-function or the pcf should be preferred in all cases.
The test statistic POINT cannot be recommended for the pcf. For the other two second-order characteristics $K$ and $L$, POINT should only be used if a clear interaction distance is known a priori as the range with high power is considerably smaller than for the other three test statistics. For $r_{\max}$ larger than this interaction distance, POINT is not able to distinguish the alternatives from CSR.
All these observations agree with the general recommendations of \citet{baddeley2014, spatstatBuch} that the upper bound should be chosen slightly larger than the interaction distance of the point process when using second-order characteristics.

For the remaining three classical summary statistics $F$, $G$ and $J$, MAD is often slightly more powerful than DCLF and INT, but also more sensitive to the choice of the upper bound $r_{\max}$. The same conclusion can also be drawn for the other three observation windows with the results being available in the online repository.
The POINT evaluation test statistic is powerful and competitive in combination with $F$, $G$ and $J$. In most cases, the power is similar to that of the MAD test statistic, with the same drawback of the sensitivity w.r.t. the choice of the specific evaluation point, which is chosen as the upper bound $r_{\max}$ of the interval used for MAD. 

For the topological summary statistics in Figure~\ref{fig:dclf-mad-int} we can observe that the DCLF test statistic always yields more powerful tests compared to the MAD test statistic for the null hypothesis of CSR. The INT test statistic can be very powerful and even outperform DCLF but the power depends extremely on the upper bound $r_{\max}$. The dependence on $r_{\max}$ is even more severe for the POINT test statistic. For the three integer-valued summary statistics $\beta_0, \beta_1$ and $\chi$, the power fluctuates a lot, and we obtain competitive tests only at a few selected distances. The use of POINT with these topological summary statistics is therefore discouraged. For the two cumulative and hence real-valued functions $\operatorname{APF}_0$ and $\operatorname{APF}_1$, POINT behaves similarly to MAD while also picking up the fluctuations of the INT test statistic.

Except for $\beta_0$, we notice a clear drop in power at some distances when using the topological summary statistics with DCLF, MAD and INT. This can be explained by comparing the pointwise means and variances of the functional summary statistics shown in Figure~\ref{fig:tda-means}. 
The $1$-dimensional Betti curves of the five alternative models differ the most from the one of the Poi null model in the location and shape of the peak. Thus, the tests are most powerful when they use the entire information available up to the peak. Similar to the second-order characteristics, the power drops slightly when choosing an upper bound that is even larger than the distance at which the peak occurs. 
For the two accumulated persistence functions, we notice large pointwise variances under the null model. The curves from different models overlap and cross each other, leading to decreases in power, in particular when using the INT test statistic. In general, we recommend using the largest possible upper bound for these two functional summary statistics in order to take the entire persistent homology information into account.
The Euler characteristic $\chi$ is the most powerful at small distances, where almost no $1$-dimensional features exist. The characteristic is negative if there are more $1$-dimensional than $0$-dimensional features, which is the case around the peak of $\beta_1$. Hence, the power with INT decreases at the distance where $\beta_1$ with INT is the most powerful. Increasing the upper bound for $\chi$ does not recover the highest power for all models since all curves eventually converge to the same value of $1$. Consequently, using INT with $\chi$ is discouraged and we recommend using DCLF instead.

In general, for testing CSR, topological summary statistics using $0$-dimensional features ($\beta_0, \operatorname{APF}_0, \chi$) are more powerful than the $1$-dimensional ones. The $1$-dimensional characteristics that consider circular ``holes'' typically require longer distances to be able to discriminate between CSR and any dependence between the points. 


\subsubsection*{Test statistics QDIR, ST and FUN}

\begin{figure}[th]
    \centering
    \subfloat{\includegraphics[width=0.9\linewidth]{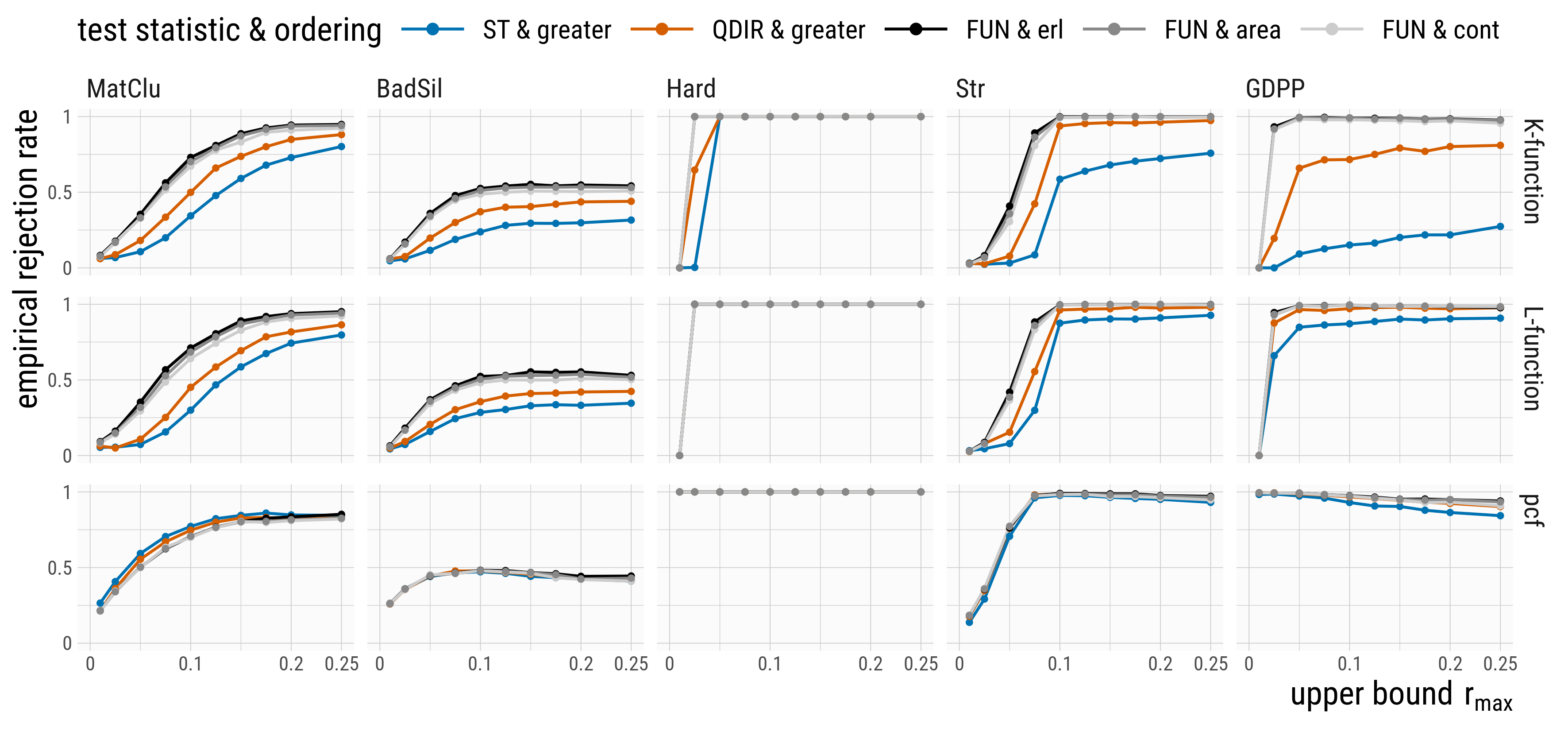}}
    \vspace{-4ex}
    \subfloat{\includegraphics[width=0.9\linewidth]{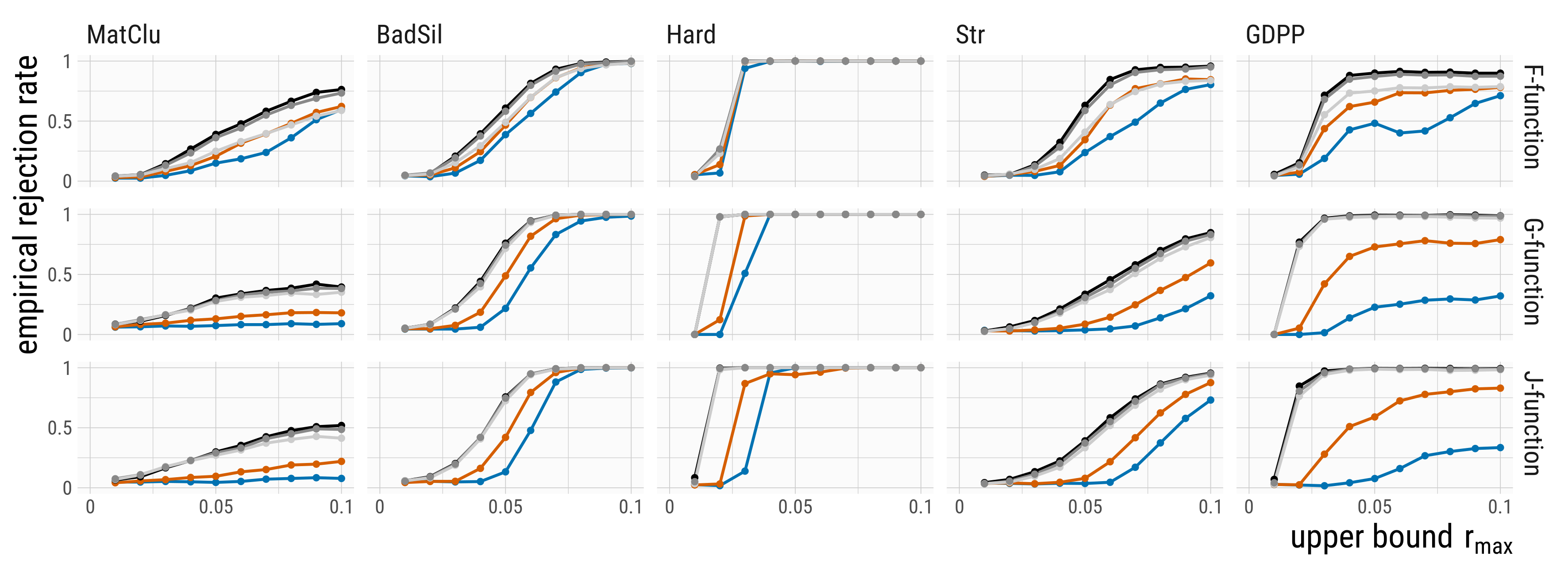}}
    \vspace{-4ex}
    \subfloat{\includegraphics[width=0.9\linewidth]{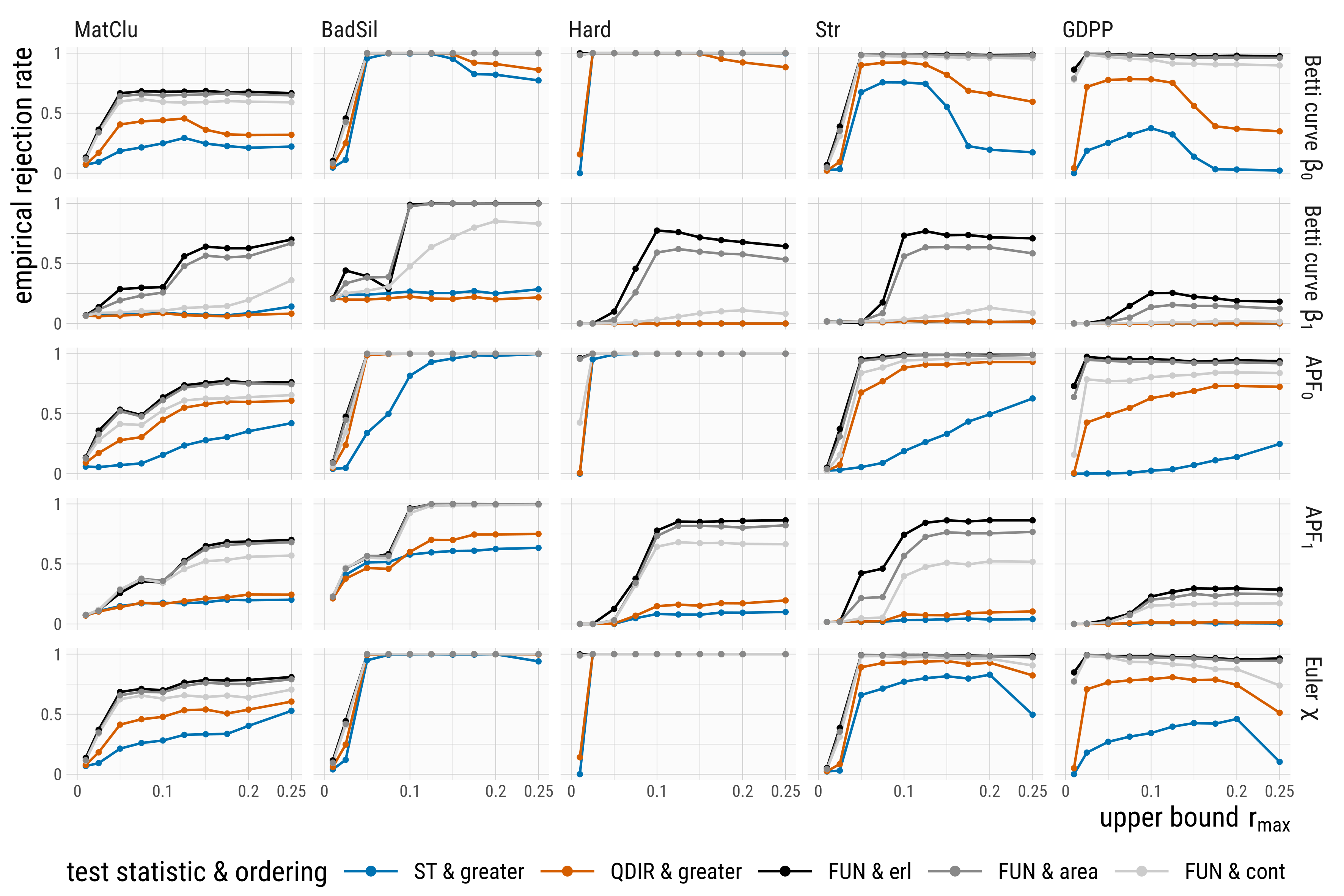}}

    \caption{Empirical power curves for the test statistics ST, QDIR and FUN with respect to the upper bound $r_{\max}$ of the domain $\mathcal{R}^*$. We used $m=299$ simulations on the observation window $W_6$.}
    \label{fig:qdir-fun}
\end{figure}

The empirical power curves for the FUN test statistic with the three types of ordering and the scaled MAD test statistics ST and QDIR are shown in Figure~\ref{fig:qdir-fun}.

Consider first the FUN test statistic with the three types of ordering. For all five models and the classical second-order summary statistics $K$, $L$ and pcf as well as $G$ and $J$, we obtain almost the same empirical rejection rates for all three orderings. For the empty space function $F$, the cont ordering yields less powerful tests than the other two orderings.

For the topological summary statistics, the difference between the three orderings becomes clearer. In particular, erl gives the most powerful tests, followed by area and then cont. The continuous rank ordering performs considerably worse than the other two if used with the $1$-dimensional Betti curve $\beta_1$. We provide in Appendix~\ref{sec:cont-details} an illustrative example that shows why the cont ordering is not a good choice for this specific summary statistic.

Overall in the setting of testing CSR, the erl ordering can be recommended for all the considered functional summary statistics.

Next, we concentrate on the two scaling approaches, ST and QDIR. The figure clearly shows that QDIR is significantly more powerful than ST. The only exception is the pair correlation function, where the choice between the five test statistics has no influence, and the purely $1$-dimensional topological statistics $\beta_1$ and $\operatorname{APF}_1$ where both scaling approaches do not work at all. 

In case of the topological summary statistics, recall that $\beta_0$, $\beta_1$ and $\chi$ are integer-valued. At each evaluation point, we have a highly discrete distribution of a small number of possible values. Thus the denominators of the test statistics ST and QDIR given as the empirical standard deviation and the absolute difference between an empirical quantile and the sample mean, respectively, can be very small.  This results in very large test statistics even for the simulations of the null model. Consequently the tests are less powerful. The problem is especially severe for $\beta_1$ at small distances. The accumulated persistence functions, in particular $\operatorname{APF}_1$, inherit the problem of having only few possible values from the corresponding Betti numbers.

\begin{table*}[th]\centering
\renewcommand{\arraystretch}{1.5}
\rowcolors{3}{gray!10}{white}
\begin{tabularx}{\textwidth}{Lcccccccccc} \toprule
\hiderowcolors 
\multirow{2}{*}{\shortstack[l]{Functional summary \\ statistic $T$}}& \multicolumn{10}{c}{upper bound $r_{\max}$} \\
\cmidrule(l){2-11}
 & 0.01 & 0.025 & 0.05 & 0.075 & 0.1 & 0.125 & 0.15 & 0.175 & 0.2 & 0.25\\
\cmidrule(l){2-11}
\showrowcolors
$\beta_0$ & 0.052 & 0.053 & 0.057 & 0.052 & 0.059 & 0.048 & 0.049 & 0.056 & 0.062 & 0.058\\
$\beta_1$ & \textbf{0.036} & 0.041 & 0.046 & \textbf{0.038} & \textbf{0.038} & \textbf{0.037} & \textbf{0.035} & \textbf{0.032} & \textbf{0.036} & \textbf{0.037}\\
$\operatorname{APF}_0$ & 0.052 & 0.053 & 0.060 & 0.050 & 0.060 & 0.052 & 0.058 & 0.058 & 0.058 & 0.062\\
$\operatorname{APF}_1$ & 0.039 & 0.051 & 0.056 & 0.051 & 0.048 & 0.050 & 0.054 & 0.052 & 0.049 & 0.051\\
$\chi$ & 0.051 & 0.058 & 0.063 & 0.060 & 0.062 & 0.060 & 0.057 & \textbf{0.067} & 0.065 & 0.042\\
 \bottomrule
\end{tabularx}
\caption{Empirical rejection rates for the nominal level $\alpha=0.05$ for the test with test statistic QDIR obtained for the CSR-model Poi from $1000$ realizations on $W_6$. Highlighted in bold are the empirical rates where the approximated $0.95$-CI for the true $p$-value does not contain the nominal level.}
\label{tab:qdir}
\end{table*}

The problem of combining topological summary statistics with scaling approaches is also illustrated in the empirical sizes of the tests obtained with QDIR, which are listed in Table~\ref{tab:qdir}. The tests with QDIR and $\beta_1$ are conservative based on the results obtained from $1000$ replications of the test. The empirical sizes of ST are similar. 

From all results shown in Figure~\ref{fig:qdir-fun} it is clear that when using the same number of realizations $m$, the FUN test statistic should be preferred over the scaling approaches. Not only does it provide more powerful tests than QDIR, but it is also more robust with respect to the upper bound $r_{\max}$. 

\subsubsection*{Test statistic CRPS}

\begin{figure}[th]
    \centering
    \subfloat{\includegraphics[width=0.9\linewidth]{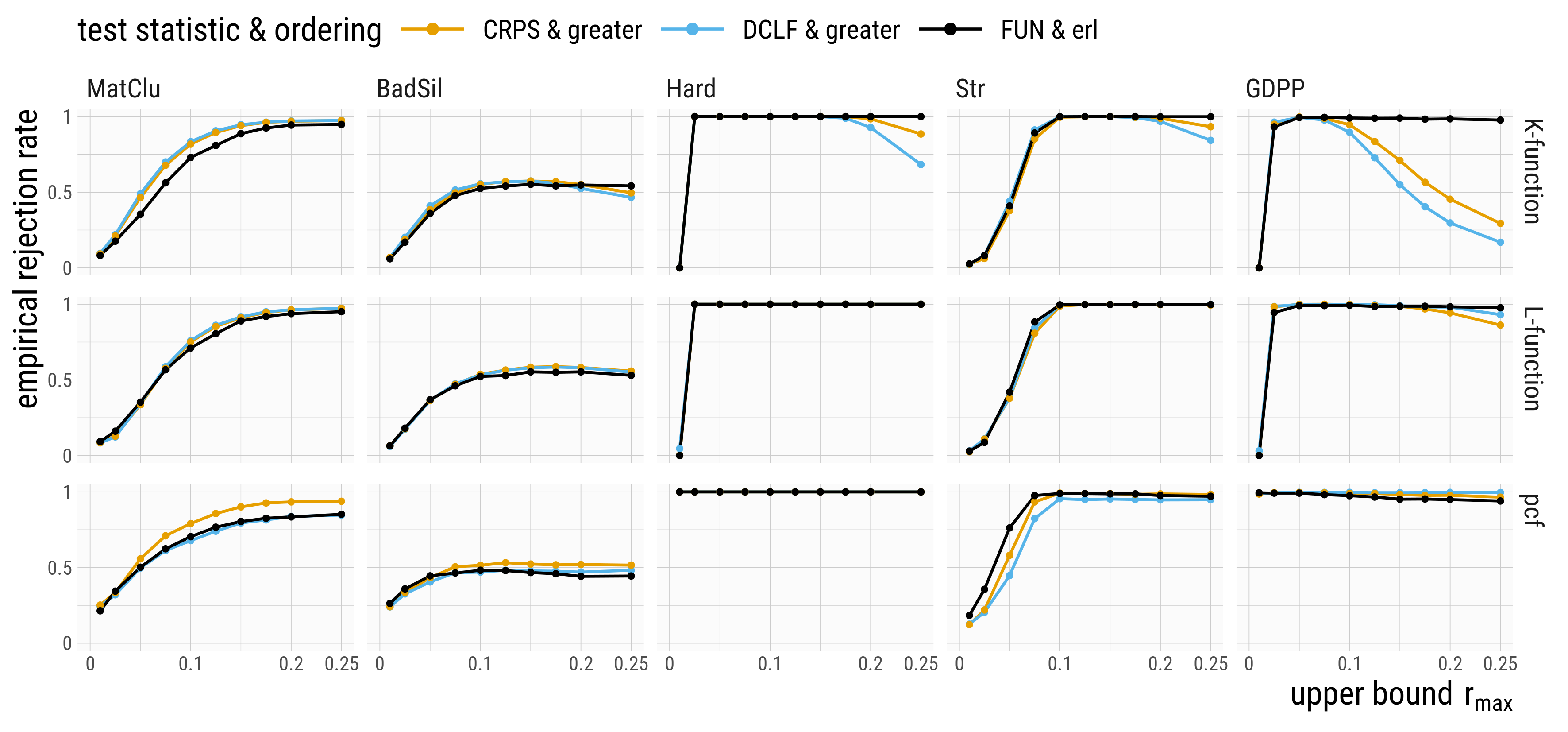}}
    \vspace{-4ex}
    \subfloat{\includegraphics[width=0.9\linewidth]{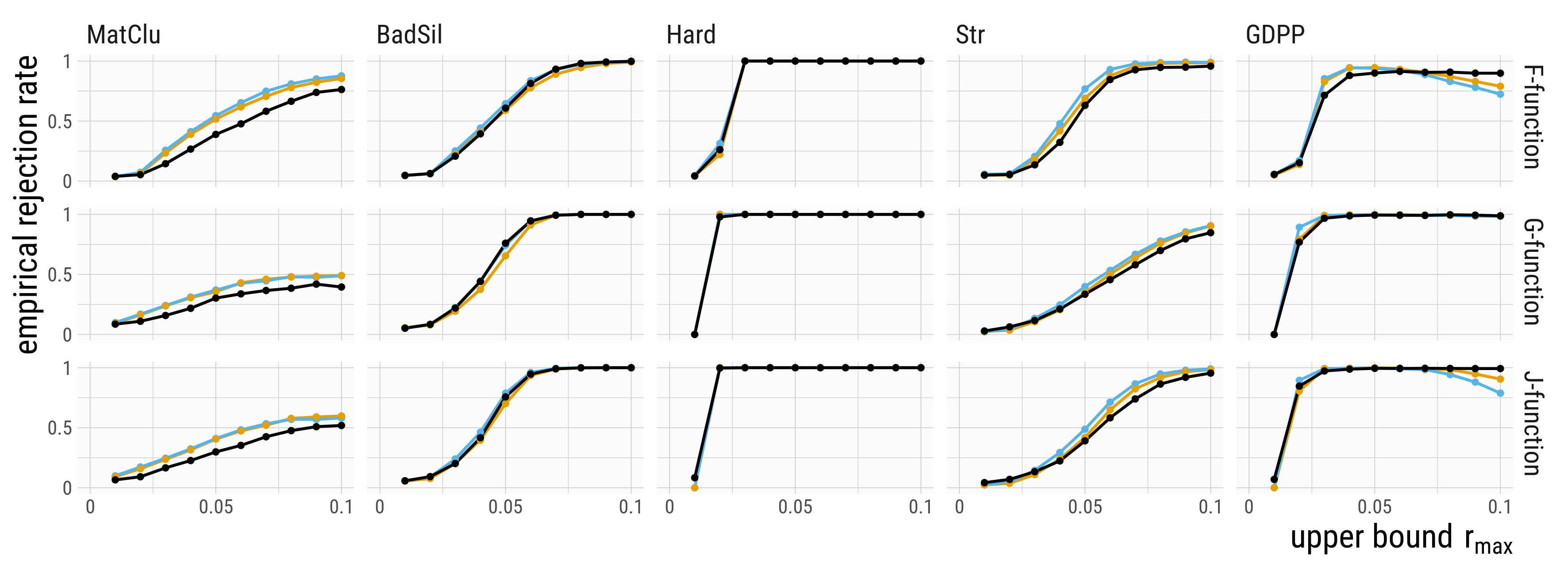}}
    \vspace{-4ex}
    \subfloat{\includegraphics[width=0.9\linewidth]{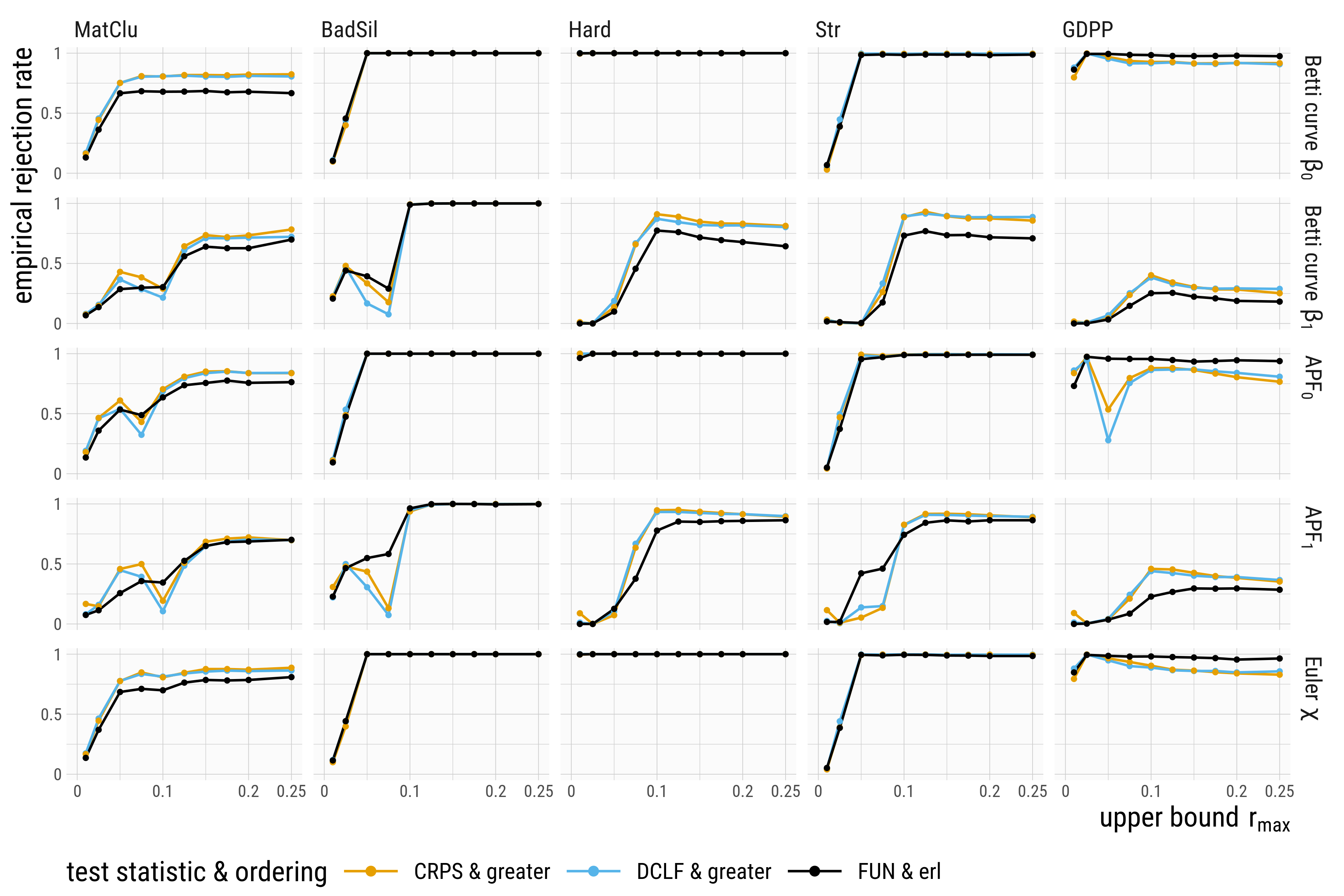}}
    
    \caption{Empirical power curves for the test statistics CRPS, DCLF and FUN with the erl ordering with respect to the upper bound $r_{\max}$ of the domain $\mathcal{R}^*$. We used $m=299$ simulations on the observation window $W_6$.}
    \label{fig:crps}
\end{figure}

Finally, we compare DCLF, FUN with erl and the recently introduced CRPS test statistic. The corresponding power curves are shown in Figure~\ref{fig:crps}.

Overall, the DCLF and CRPS test statistics result in tests with comparable power for most functional summary statistics. This makes sense due to the construction as both consider in some sense the squared deviation to the null model.
In cases where the variability of the functional summary statistics under the null model is high, CRPS is more powerful. For example, this is the case for the $K$- and $J$-function at larger distances, the pair correlation function, and the $1$-dimensional topological characteristics $\beta_1$ and $\operatorname{APF}_1$. 

Compared with the two deviation-based test statistics DCLF and CRPS, FUN with erl results in more robust tests with respect to the varying upper bound $r_{\max}$ not only for the classical summary statistics, but also for the topological ones. As mentioned above, the drop in power for $\beta_1$ and $\operatorname{APF}_1$ with deviation-based test statistics is noticeable for most alternatives and appears even on the largest window $W_{20}$ with about $1000$ points. With FUN and the erl ordering, we do not observe any drop. 

When considering the absolute power for alternatives that exhibit clustering as MatClu, we can conclude that deviation-based test statistics are more powerful than FUN. For the repulsive GDPP model, the FUN test statistic with erl ordering is more powerful than DCLF and CRPS for all functional summary statistics except for the purely $1$-dimensional topological summary statistics $\beta_1$ and $\operatorname{APF}_1$.

\subsubsection*{Number of simulations $m$}

The remaining open question regarding Monte Carlo tests with a single functional summary statistic is the number of simulations $m$. So far, we discussed the results obtained using $m=299$ for all test statistics. Although this choice is larger than the numbers often used for the deviation-based test statistics of type A in the literature, it is smaller than recommendations for the FUN test statistic. 

\begin{figure}[th]
    \centering
    \subfloat{\includegraphics[width=0.9\linewidth]{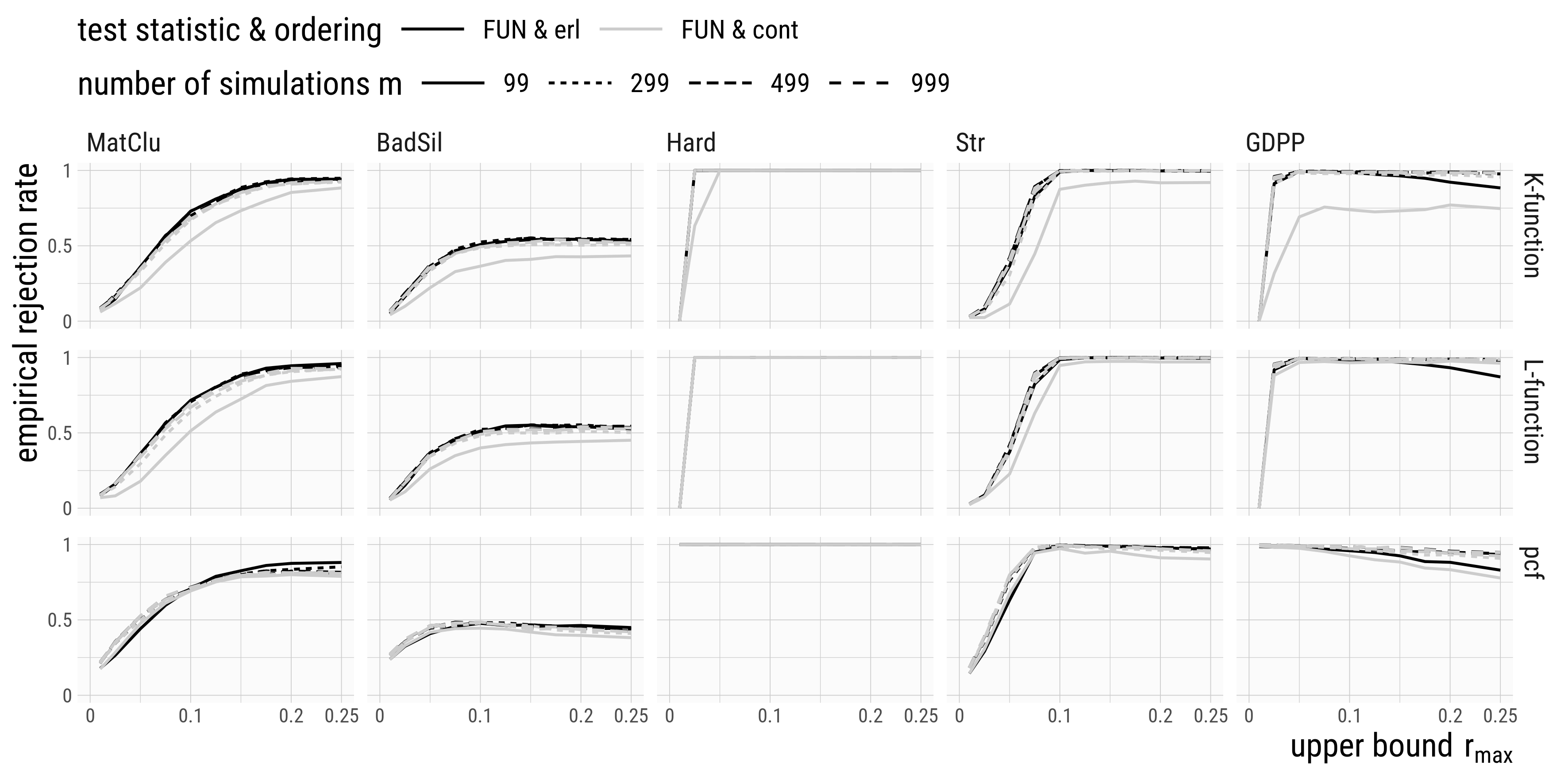}}
    \vspace{-4ex}
    \subfloat{\includegraphics[width=0.9\linewidth]{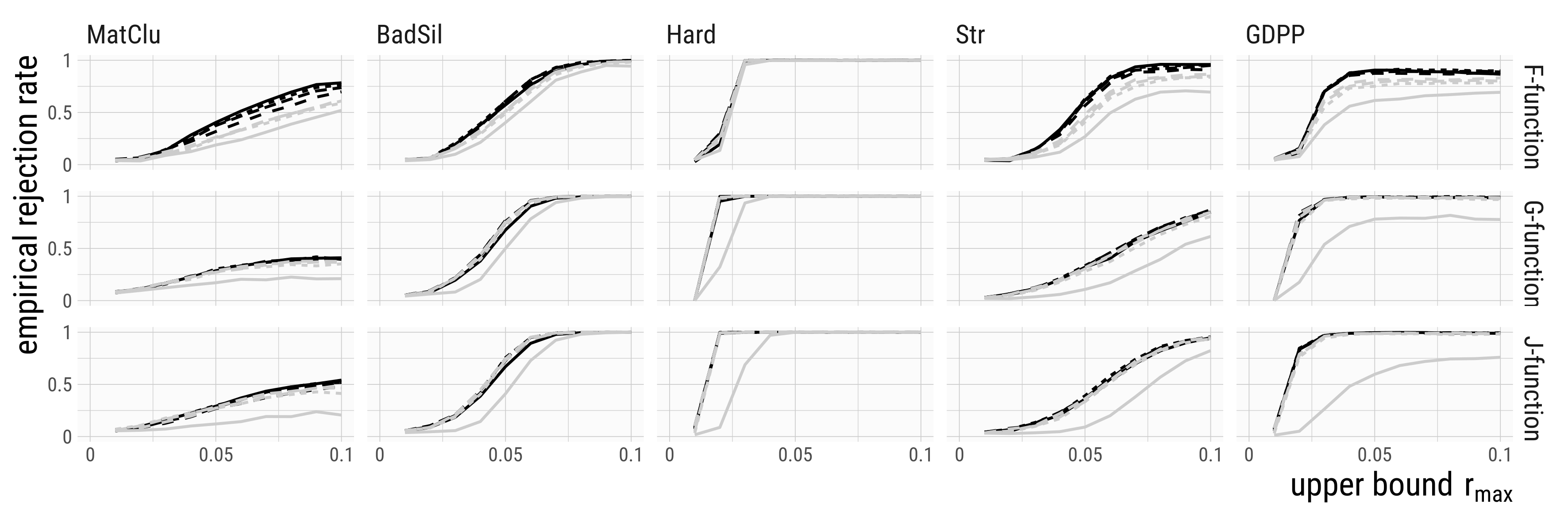}}
    \vspace{-4ex}
    \subfloat{\includegraphics[width=0.9\linewidth]{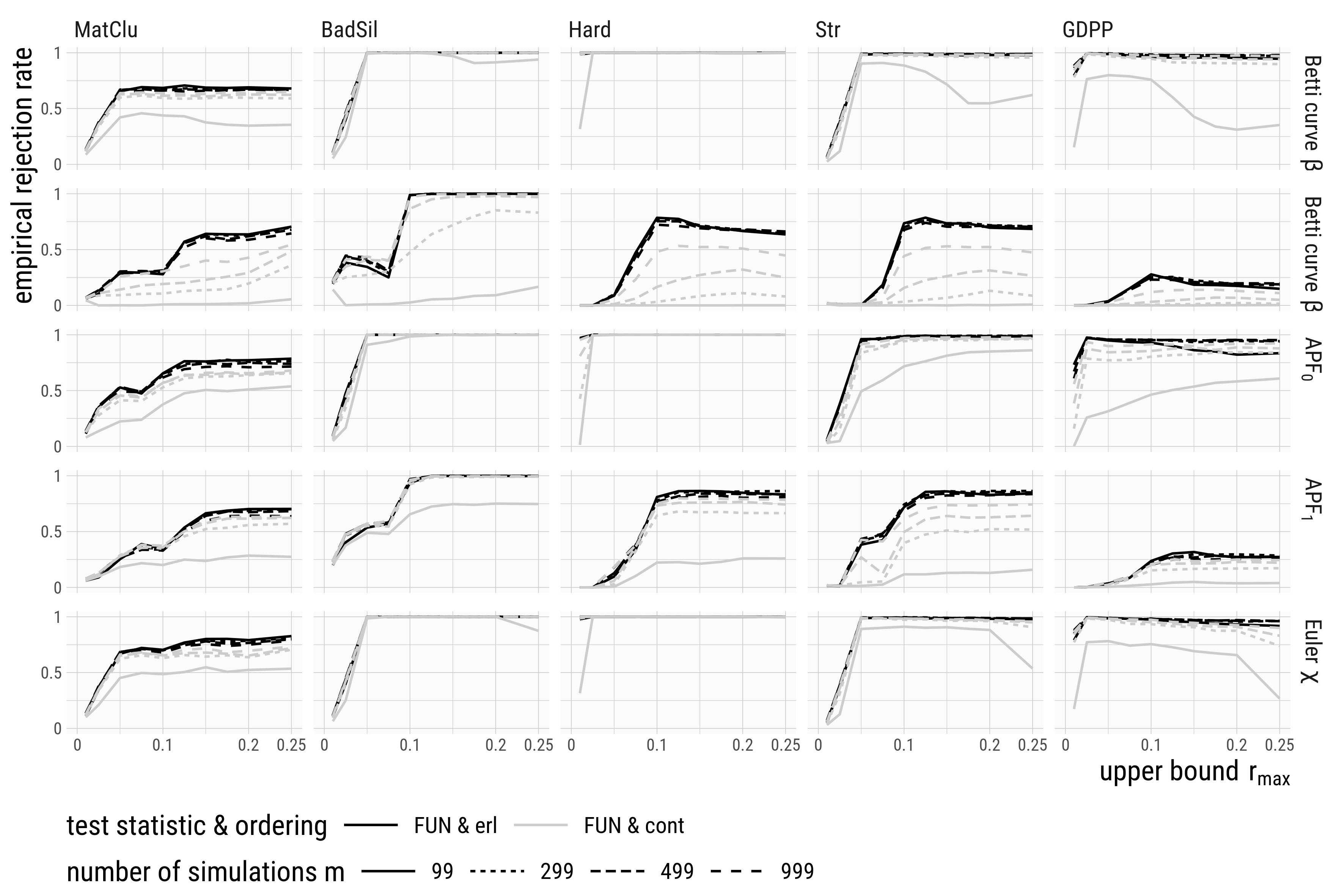}}
    
    \caption{Empirical power curves for the test statistics FUN with either the erl or the cont ordering and varying number of simulations $m$ on the observation window $W_6$.}
    \label{fig:m-fun}
\end{figure}

Figure~\ref{fig:m-fun} displays the rejection rate curves obtained for the FUN test statistic with either the erl or the cont ordering for different choices of the number of simulations $m$. As discussed previously, the continuous rank measure is less powerful than erl in our setting. We can also see that cont with $m=999$ is often still less powerful than erl with $m=99$ simulations, while being computationally much more expensive. The only big increase of power due to a higher number of simulations can be observed when going from $m=99$ to $m=299$ in case of the cont measure. This happens both for the classical and the topological functional summary statistics. 

Due to the aforementioned problems with the $1$-dimensional topological features, cont benefits from a higher number of simulations for that specific summary statistics. Hence, only for $\beta_1$ and $\operatorname{APF}_1$ an ongoing increase in power is visible. There are only minor differences for the erl ordering in our study. It should be noted that for several functional summary statistics either $m=99$ or $m=299$ yields the most powerful test, and with $m=999$ the power was considerably lower.
Based on this simulation study, there is no evidence to recommend a higher number than $m=299$ for the simulations when using the FUN test statistic with erl ordering.

\begin{figure}[th]
    \centering
    \subfloat{\includegraphics[width=0.9\linewidth]{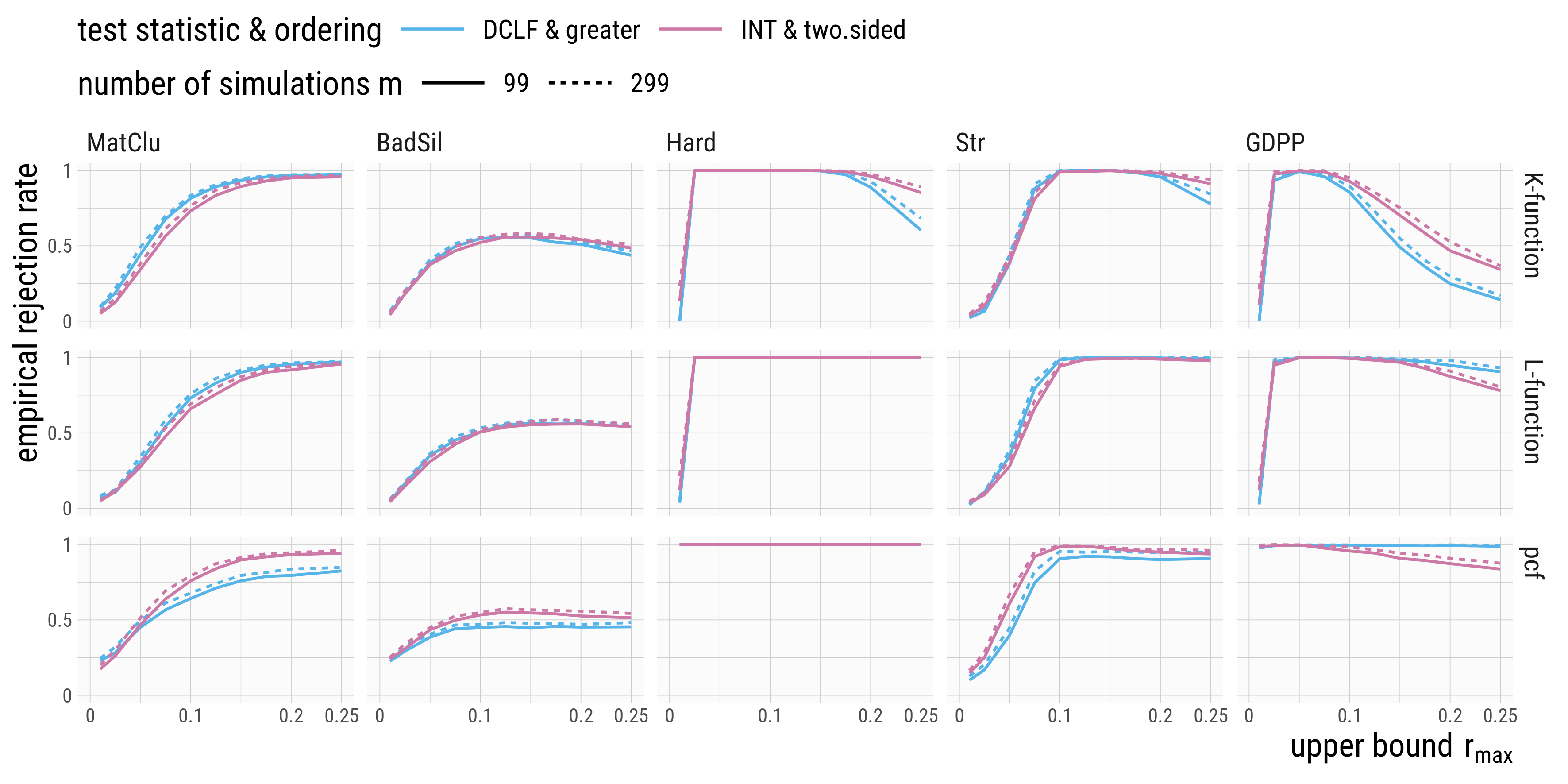}}
    \vspace{-4ex}
    \subfloat{\includegraphics[width=0.9\linewidth]{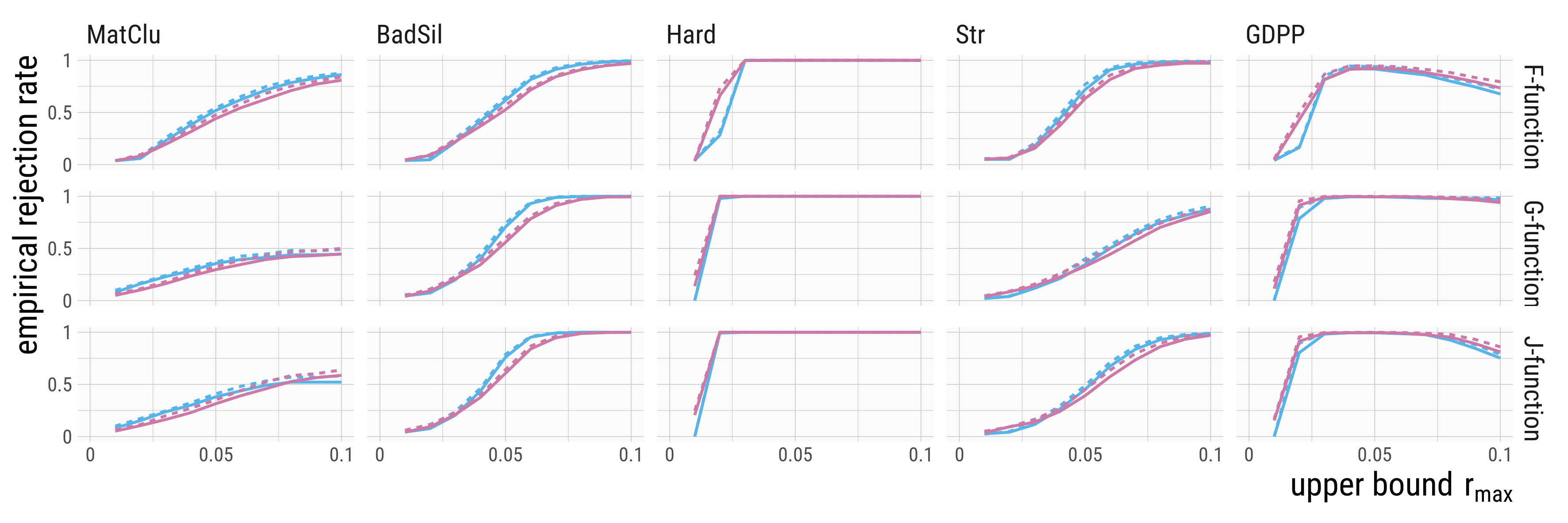}}
    \vspace{-4ex}
    \subfloat{\includegraphics[width=0.9\linewidth]{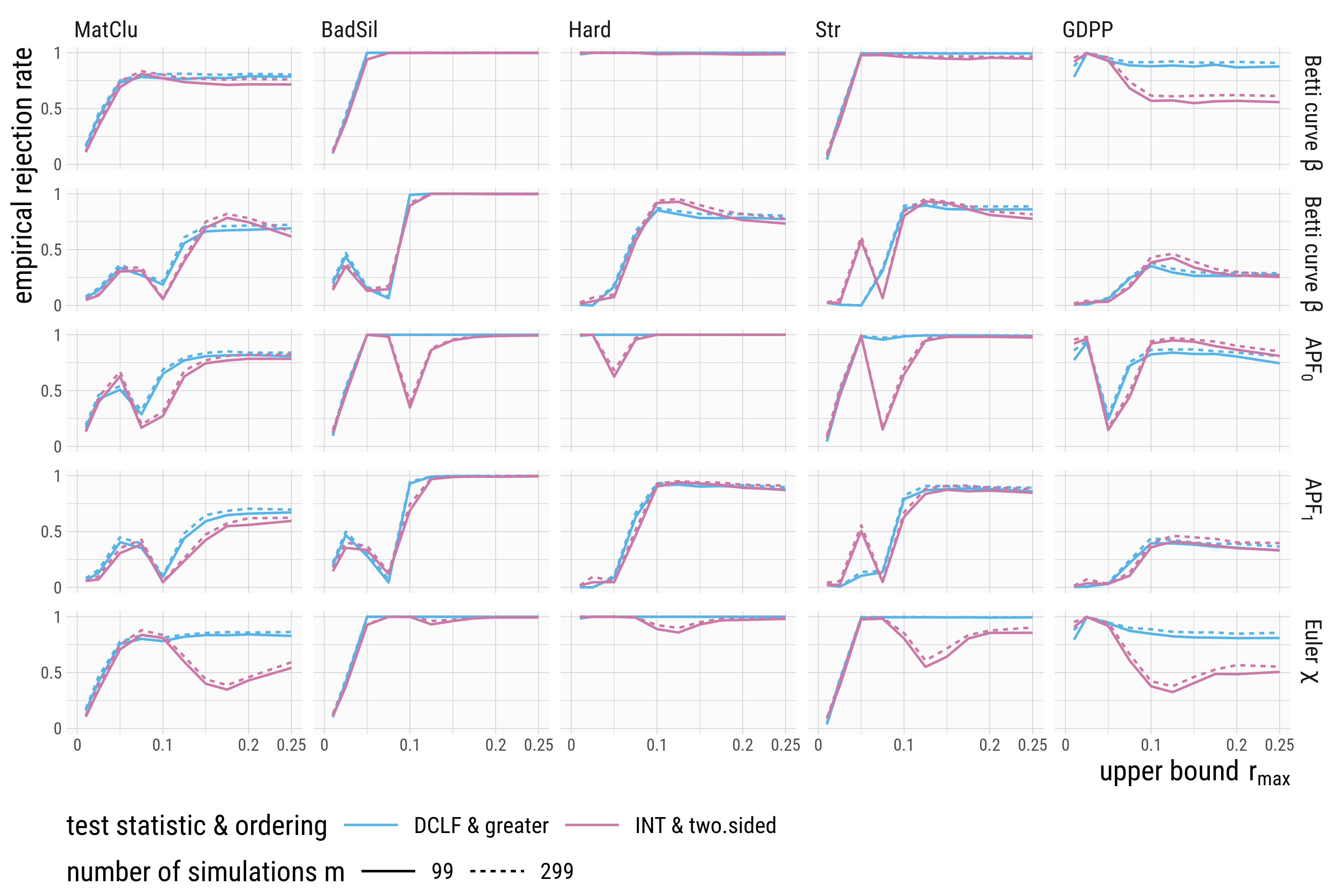}}
    
    \caption{Empirical power curves with respect to the upper bound $r_{\max}$ of the domain $\mathcal{R}^*$ with observation window $W_6$.}
    \label{fig:m-other}
\end{figure}

As mentioned in \citet{fend2025}, many authors use $m=99$ for MAD or DCLF. In our results shown in Figure~\ref{fig:m-other} the power curves for DCLF and INT for $m=299$ are compared with those obtained when using $m=99$. Overall, the differences are rather small on the observation window $W_6$. On the two smaller observation windows, the increase from using $m=299$ is larger and thus if it is computationally feasible we recommend $m=299$ for test statistics of type A and B.

\subsection{Combination of several functional summary statistics}\label{sec:combi-res}

\begin{table*}[th]\centering
\renewcommand{\arraystretch}{1.7}
\rowcolors{1}{gray!10}{white}
\begin{tabularx}{\linewidth}{lL@{\hskip 15pt}cccc@{\hskip 15pt}cccccc@{\hskip 15pt}c}
\toprule
\hiderowcolors 
& Model & $L$ & $J$ & $\beta_0$ & $\chi$ & $\beta_0$, $L$ & $\chi$, $L$ & $\beta_0$, $J$ & $\chi$, $J$ & $J$, $L$ & $\chi$, $\beta_0$ & $\beta_0$, $J$, $L$\\
\showrowcolors
\midrule
\cellcolor{white} & MatClu & \textbf{0.274} & 0.122 & 0.158 & 0.181 & 0.256 & \textbf{0.288} & 0.144 & 0.173 & 0.278 & 0.193 & 0.270\\
 \cellcolor{white}& BadSil & \textbf{0.746} & 0.592 & 0.645 & 0.653 & 0.790 & 0.775 & 0.690 & 0.673 & 0.807 & 0.645 & \textbf{0.822}\\
 \cellcolor{white}& Hard & 0.902 & \textbf{0.968} & 0.948 & 0.871 & 0.899 & 0.809 & \textbf{0.956} & 0.945 & 0.945 & 0.928 & 0.944\\
 \cellcolor{white}& Str & \textbf{0.332} & 0.216 & 0.277 & 0.248 & 0.286 & 0.270 & \textbf{0.295} & 0.242 & \textbf{0.295} & 0.253 & 0.290\\
\parbox[t]{2mm}{\multirow{-5}{*}{\cellcolor{white}\rotatebox[origin=c]{90}{$W_1$ ($50$ points)}}}  & GDPP & 0.168 & \textbf{0.269} & 0.178 & 0.122 & 0.140 & 0.121 & 0.222 & 0.221 & \textbf{0.232} & 0.140 & 0.196\\ \cmidrule{2-13}
\cellcolor{gray!10} & MatClu & \textbf{0.535} & 0.198 & 0.298 & 0.375 & 0.527 & \textbf{0.546} & 0.286 & 0.342 & 0.514 & 0.366 & 0.504\\
\cellcolor{gray!10} & BadSil & 0.550 & 0.924 & 0.960 & \textbf{0.966} & 0.978 & 0.974 & 0.982 & 0.977 & 0.961 & 0.965 & \textbf{0.986}\\
\cellcolor{gray!10} & Hard & 1.000 & 1.000 & 1.000 & 1.000 & 1.000 & 0.999 & 1.000 & 1.000 & 1.000 & 1.000 & 1.000\\
\cellcolor{gray!10} & Str & \textbf{0.683} & 0.454 & 0.568 & 0.523 & 0.642 & 0.627 & 0.563 & 0.521 & \textbf{0.646} & 0.532 & 0.607\\
\parbox[t]{2mm}{\multirow{-5}{*}{\cellcolor{gray!10}\rotatebox[origin=c]{90}{$W_2$ ($100$ points)}}} & GDPP & 0.406 & \textbf{0.566} & 0.426 & 0.349 & 0.350 & 0.321 & \textbf{0.505} & 0.476 & 0.483 & 0.367 & 0.433\\ \cmidrule{2-13} 
\cellcolor{white} & MatClu & \textbf{0.951} & 0.519 & 0.679 & 0.809 & \textbf{0.944} & \textbf{0.944} & 0.649 & 0.782 & \textbf{0.944} & 0.785 & 0.936\\
\cellcolor{white} & BadSil & 0.530 & 1.000 & 1.000 & 1.000 & 1.000 & 1.000 & 1.000 & 1.000 & 1.000 & 1.000 & 1.000\\
\cellcolor{white} & Hard & 1.000 & 1.000 & 1.000 & 1.000 & 1.000 & 1.000 & 1.000 & 1.000 & 1.000 & 1.000 & 1.000\\
\cellcolor{white} & Str & \textbf{0.998} & 0.955 & 0.986 & 0.985 & 0.997 & \textbf{0.998} & 0.987 & 0.985 & 0.997 & 0.985 & 0.993\\
\parbox[t]{2mm}{\multirow{-5}{*}{\cellcolor{white}\rotatebox[origin=c]{90}{$W_6$ ($300$ points)}}} & GDPP & 0.977 & \textbf{0.993} & 0.984 & 0.964 & 0.959 & 0.957 & \textbf{0.987} & \textbf{0.987} & 0.985 & 0.969 & 0.981\\ \cmidrule{2-13}
\cellcolor{gray!10} & MatClu & 1.000 & 0.982 & 0.996 & 1.000 & 1.000 & 1.000 & 0.994 & 1.000 & 1.000 & 0.999 & 1.000\\
\cellcolor{gray!10} & BadSil & 0.546 & 1.000 & 1.000 & 1.000 & 1.000 & 1.000 & 1.000 & 1.000 & 1.000 & 1.000 & 1.000\\
\cellcolor{gray!10} & Hard & 1.000 & 1.000 & 1.000 & 1.000 & 1.000 & 1.000 & 1.000 & 1.000 & 1.000 & 1.000 & 1.000\\
\cellcolor{gray!10} & Str & 1.000 & 1.000 & 1.000 & 1.000 & 1.000 & 1.000 & 1.000 & 1.000 & 1.000 & 1.000 & 1.000\\
\parbox[t]{2mm}{\multirow{-5}{*}{\cellcolor{gray!10}\rotatebox[origin=c]{90}{$W_{20}$ ($1000$ points)}}} & GDPP & 1.000 & 1.000 & 1.000 & 1.000 & 1.000 & 1.000 & 1.000 & 1.000 & 1.000 & 1.000 & 1.000\\
\bottomrule
\end{tabularx}
\caption{Empirical power estimates from $1000$ realizations when using the test statistic FUN and the ordering erl. The function parameters are stated in the text. For each row, the bold values indicate the highest power from a single summary function and from a combination. Rows with more than two perfect rejection rates of $1.000$ per type (single or combination) are not printed in bold face.}
\label{tab:res-combi}
\end{table*}

We compared tests that use a single summary statistic with similar tests that combine several of them. From the results obtained in the previous experiments, we selected the $L$-function as best second-order characteristic, the $J$-function representing the distance-based summaries, and the $0$-dimensional Betti curve and the Euler characteristic curve as topological summary statistics. We selected two topological summary functions as $\beta_0$ only takes the information on the $0$-dimensional topological features into account while $\chi$ also includes the $1$-dimensional features. For the comparison we used the FUN test statistic with the erl ordering where we discretized each functional summary statistic at $513$ equidistant evaluation points. In all tests, we used $m=299$ simulations from CSR. The combined tests were conducted using the two-step combination procedure available in the \texttt{GET}-function \texttt{global\_envelope\_test} \citep{GET}. 

The $L$-function was estimated on the interval $[0,0.25]$, the $J$-function on $[0,0.1]$, the $0$-dimensional Betti curve $\beta_0$ on $[0,0.1]$ and the Euler characteristic curve $\chi$ on $[0,0.25]$. The selected upper bounds were chosen as those that result in the highest overall powers for the individual summary functions. Note that this choice is not necessarily optimal for all of the non-CSR point process models. As an example, the Euler characteristic curve $\chi$ is more powerful against a GDPP alternative at very small distances up to $0.03$.

In Table~\ref{tab:res-combi} the empirical power estimates for the compared tests are listed. The empirical rejection rates for the CSR-model Poi were in all cases close to the nominal level of $\alpha=0.05$.

From the estimates, we can conclude that the combined tests are often similar in power to the most powerful individual test. This aligns with the results in \citet{mrkvicka2017} regarding combinations of the classical summary functions. 

When used individually, the topological summary functions are rarely more powerful than the classical summary functions. But in combination with either of the two classical summary functions, we obtain tests that can be even more powerful than combining $L$ and $J$. This is, for example, the case for the Baddeley-Silverman cell process on the two smaller windows. A combination of at least one of the classical summary functions with a topological one led to a power that is higher than the individual powers and higher than the combination of $L$ and $J$. Another example is the combination of $\chi$ and $L$ for the clustered alternative MatClu. For more regular alternatives, $\beta_0$ and $J$ turned out to be a powerful combination in our experiments. 

The empirical power of the combination of the three functions $\beta_0, J$ and $L$ is, except for the Baddeley-Silverman cell process, often slightly smaller than the power for a combination of two summary functions. In these cases, adding the third type of information is not beneficial.

\section{Discussion} \label{sec:discussion}

The focus of this study was on comparing the performance of classical and topological functional summary statistics in a common goodness-of-fit testing scenario. Our common test setting is the null hypothesis of complete spatial randomness of a planar point process with five stationary and isotropic alternatives that deviate from CSR in different ways. We investigated the empirical sizes and powers of a diverse set of tests. In particular different constructions for the test statistic as well as different orderings for vector-valued test statistics were included. Several tuning parameters such as the number of simulations in the Monte Carlo test and the relevant part of the domain of the functional summary statistic affect the performance of the test. This is the reason why we investigated their influence to see which of the tests are more robust when these parameters are adapted. With many possible parameters taken into account, we obtained a large number of distinct specific tests that were conducted. Additionally, we considered four different sizes for the observation window. The results for $W_6$ with approximately $300$ points per point pattern are shown in \Cref{fig:dclf-mad-int,fig:qdir-fun,fig:crps,fig:m-fun,fig:m-other} and the results for $W_1$, $W_2$ and $W_{20}$ are available in our online repository.

\begin{table*}\centering
\begin{threeparttable}
\renewcommand{\arraystretch}{1.5}
\setlength{\tabcolsep}{3pt}
\rowcolors{3}{gray!15}{gray!5}
\begin{tabularx}{\textwidth}{>{\,}llCCCCCCCCCCC} \toprule
\hiderowcolors 
\multicolumn{2}{l}{\multirow{3}{*}{\textbf{Test Statistic} $D$}} &  \multicolumn{11}{c}{\textbf{Functional Summary Statistic} $T$} \\  \cmidrule{3-13}
& & \multicolumn{3}{c}{Second order} &  \multicolumn{3}{c}{Distance-based} & \multicolumn{5}{c}{TDA-based}\\
\cmidrule(r){3-5} \cmidrule(r){6-8} \cmidrule{9-13}
& & $K$ & $L$ & $pcf$ & $F$ & $G$ & $J$ & $\beta_0$ & $\beta_1$ & $\operatorname{APF}_0$ & $\operatorname{APF}_1$ & $\chi$\\ \midrule
\multicolumn{2}{l}{\rowgroup{\textbf{Type A}}}\\ \cmidrule{1-2}
\showrowcolors
\multicolumn{2}{l}{\cellcolor{white}$\operatorname{MAD}$} & (\cmark) & (\cmark) & \xmark & \cmark\tnote{\bestmark} & \cmark & \cmark & \cmark & \xmark & \xmark   & \xmark   &  \xmark \\
\multicolumn{2}{l}{\cellcolor{white}$\operatorname{DCLF}$} & (\cmark) & \cmark\tnote{\bestmark}  & \cmark & \cmark & \cmark\tnote{\bestmark}  & \cmark &\cmark\tnote{\bestmark}  & \cmark & \cmark & \cmark & \cmark\tnote{\bestmark}  \\
\multicolumn{2}{l}{\cellcolor{white}$\operatorname{QDIR}$} & \cmark & (\cmark) & \cmark & \xmark & \xmark & \xmark &\xmark  &\xmark  &\xmark  & \xmark & \xmark \\[0.75ex]
\multicolumn{2}{l}{\cellcolor{white}$\operatorname{ST}$} & \xmark & \xmark & \cmark & \xmark & \xmark & \xmark & \xmark &\xmark &\xmark &\xmark &\xmark   \\[0.75ex]
\multicolumn{2}{l}{\cellcolor{white}$\operatorname{CRPS}$} & (\cmark) & \cmark & \cmark\tnote{\bestmark} & \cmark & \cmark & \cmark\tnote{\bestmark}& \cmark& \cmark\tnote{\bestmark} & \cmark& \cmark\tnote{\bestmark}& \cmark \\[1.25ex] 
\hiderowcolors 
\multicolumn{2}{l}{\rowgroup{\textbf{Type B}}}\\ \cmidrule{1-2}
\showrowcolors
\multicolumn{2}{l}{\cellcolor{white}$\operatorname{INT}$} & (\cmark) & \cmark & \cmark\tnote{\bestmark} & \cmark & \cmark & \cmark & (\cmark)  & \cmark & (\cmark)  & (\cmark) & \xmark  \\
\multicolumn{2}{l}{\cellcolor{white}$\operatorname{POINT}$} & \xmark & \xmark & \xmark &  \cmark & (\cmark) & (\cmark) & \xmark & \xmark & \cmark & \cmark & \xmark \\[1.25ex] 
\hiderowcolors 
\multicolumn{2}{l}{\rowgroup{\textbf{Type C}}}\\ \cmidrule{1-2}
\showrowcolors
\cellcolor{white}  & \cellcolor{white}$\operatorname{erl}$ & \cmark\tnote{\bestmark} & \cmark\tnote{\bestmark}& \cmark& \cmark& \cmark\tnote{\bestmark}& \cmark\tnote{\bestmark}& \cmark\tnote{\bestmark}& \cmark& \cmark\tnote{\bestmark}& \cmark& \cmark\tnote{\bestmark}\\
\cellcolor{white}  & \cellcolor{white}$\operatorname{area}$ & \cmark & \cmark & \cmark & \cmark & \cmark & \cmark & \cmark & (\cmark) & (\cmark)& (\cmark) & \cmark \\
\multirow{-3}{0.8cm}{\cellcolor{white}$\operatorname{FUN}$  with} & \cellcolor{white}$\operatorname{cont}$ & \cmark & \cmark & \cmark & \xmark & \cmark & \cmark & \cmark & \xmark & (\cmark) & \xmark & (\cmark)   \\
\bottomrule
\end{tabularx}
\caption{General recommendations derived from our power study regarding which functional summary statistic should be used with which test statistic. Here we assume $m=299$ for all tests. Recommended combinations are indicated by \cmark, combinations which one should avoid by \xmark. In case we can recommend a test only with certain restrictions -- e.g. due to a high sensitivity w.r.t. the choice of upper bound $r_{\max}$ or due to it being slightly less powerful than another test statistic -- the symbol (\cmark) is used. For each functional summary statistic, i.e. each column, the symbol \cmark\tnote{\bestmark} is used for the overall most powerful test statistics.}
\label{tab:overall-recomm}
\end{threeparttable}
\end{table*}

The overall performance of the combination of functional summary statistic on one hand and the test statistic and ordering on the other hand is summarized in Table~\ref{tab:overall-recomm}. In particular, the test statistics DCLF, CRPS and FUN with the erl odering performed well no matter the chosen functional summary statistics. For most of the other statistics, the performance is very sensitive with respect to the tuning parameters, in particular the choice of the upper bound $r_{\max}$. Whether we can recommend such a test statistic for a specific summary function depends on how straight-forward it is to select these additional parameters. For some of the functional summary statistics, the \emph{best} test statistics depends on the type of alternative. In particular, the order of the test statistics in terms of empirical power is sometimes inverted between the clustered MatClu model and the repulsive GDPP model. In these cases, we based our recommendations on the order obtained for the majority of the alternatives.

In Section~\ref{sec:open-question} we formulated four open questions regarding goodness-of-fit testing for spatial point processes, which we are now able to answer.

The first topic addresses the performance of topological summary statistics. In contrast to the classical functional summary statistics, in particular the second-order characteristics, only selected test statistics can be recommended. For the topological summary statistics both scaling approaches QDIR and ST as well as the cont ordering perform worse than DCLF, CRPS or FUN with the erl ordering. The difference between the test statistics is the largest for the $1$-dimensional topological features, where at small distances only few holes with short lifetimes exist. This results in highly discrete pointwise distributions of the value of the functional summary statistic. The existence of such a single small hole -- that could also be considered topological noise -- can change the test decision.
Besides this, the Euler characteristic is also special, since it is the only functional summary statistic in our selection that can become negative before eventually converging to a limiting value of $1$ for increasing spatial distance. Consequently, test statistics that take absolute differences into account should be preferred for $\chi$ and the raw INT test statistic is less discriminatory.

The second open question asked for the necessary number of simulations $m$. Based on our experiments (with results shown in \Cref{fig:m-fun,fig:m-other}) we can conclude that $m=99$ is a reasonable choice for the scalar-valued test statistics of type $A$ (except QDIR and ST) and $B$. The improvement obtained from $m=299$ is more pronounced the fewer points there are in the point pattern. For the FUN test statistic $m=99$ is not enough. With the erl and area orderings, $m=299$ results in powerful tests. Any further increase sometimes even decreased the empirical power. In contrast to that, we need a higher number of simulation for the cont ordering. In many cases, even $m=999$ for cont was not as powerful as the erl ordering with $m=299$. In prior works \citep[e.g.][]{mrkvicka2017} it is stated that the scaled MAD test statistics QDIR and ST require fewer simulations than FUN with the erl ordering. In our results, QDIR and ST were rarely competitive even when using the same number of simulations.

The use of the CRPS test statistics was the third open question. By construction, DCLF and CRPS behave very similarly. We recommend CRPS over DCLF in cases where the functional summary statistic estimator has a high and non-constant pointwise variance as it resulted in more powerful tests. This is the case for the $K$-function, the pcf and the $1$-dimensional topological characteristics $\beta_1$ and $\operatorname{APF}_1$. If DCLF and CRPS are equally powerful, we recommend taking DCLF as CRPS is computationally more expensive.

Our last open question considered combinations of topological and classical functional summary statistics. Based on the results in Section~\ref{sec:combi-res} we conclude that the $L$-function and the Euler characteristic $\chi$ as well as $\beta_0$ and the $J$-function, complement each other well. The first pair is powerful in detecting clustering, whereas the second pair performs well for strongly repulsive processes. The combined tests of classical and topological summary statistics are similar or even more powerful than the individual tests, which is in agreement with the results in \citet{mrkvicka2022} obtained for combinations of the classical functional summary statistics.

In this work, we restricted ourselves to the Monte Carlo test setting. For several of the combinations, limit theorems under the CSR assumption are also available. Thus, it is still an interesting question to see how large the point pattern needs to be such that a derived asymptotic test is reasonable and powerful. Although our results provide guidelines on how to choose the test statistic and the other tuning parameters once a specific functional summary statistic has been chosen, we did not discuss how to choose the summary statistic itself. This choice in general depends a lot on the alternatives that one considers. If in doubt, the results for combined tests show that combining classical and topological functional summary statistics can yield very powerful tests that detect several different deviations from the null hypothesis. Thus, it is worthwhile to investigate these combinations in even more depth.

\bibliography{bibliography-power} 

\begin{thebibliography}{23}
\providecommand{\natexlab}[1]{#1}
\providecommand{\url}[1]{\texttt{#1}}
\expandafter\ifx\csname urlstyle\endcsname\relax
  \providecommand{\doi}[1]{doi: #1}\else
  \providecommand{\doi}{doi: \begingroup \urlstyle{rm}\Url}\fi

\bibitem[Baddeley and Turner(2005)]{spatstat}
A.~Baddeley and R.~Turner.
\newblock {spatstat}: An {R} package for analyzing spatial point patterns.
\newblock \emph{Journal of Statistical Software}, 12\penalty0 (6):\penalty0 1--42, 2005.
\newblock \doi{10.18637/jss.v012.i06}.

\bibitem[Baddeley et~al.(2014)Baddeley, Diggle, Hardegen, Lawrence, Milne, and Nair]{baddeley2014}
A.~Baddeley, P.~J. Diggle, A.~Hardegen, T.~Lawrence, R.~K. Milne, and G.~Nair.
\newblock On tests of spatial pattern based on simulation envelopes.
\newblock \emph{Ecological Monographs}, 84\penalty0 (3):\penalty0 477--489, 2014.
\newblock \doi{10.1890/13-2042.1}.

\bibitem[Baddeley et~al.(2015)Baddeley, Rubak, and Turner]{spatstatBuch}
A.~Baddeley, E.~Rubak, and R.~Turner.
\newblock \emph{Spatial Point Patterns: Methodology and Applications with {R}}.
\newblock Chapman and Hall/CRC Press, London, 2015.
\newblock ISBN 9781482210200.
\newblock URL \url{https://www.routledge.com/Spatial-Point-Patterns-Methodology-and-Applications-with-R/Baddeley-Rubak-Turner/p/book/9781482210200/}.

\bibitem[Biscio and M{\o}ller(2019)]{biscio2019}
C.~A.~N. Biscio and J.~M{\o}ller.
\newblock The accumulated persistence function, a new useful functional summary statistic for topological data analysis, with a view to brain artery trees and spatial point process applications.
\newblock \emph{Journal of Computational and Graphical Statistics}, 28\penalty0 (3):\penalty0 671--681, 2019.
\newblock \doi{10.1080/10618600.2019.1573686}.

\bibitem[Biscio et~al.(2020)Biscio, Chenavier, Hirsch, and Svane]{biscio2020}
C.~A.~N. Biscio, N.~Chenavier, C.~Hirsch, and A.~M. Svane.
\newblock Testing goodness of fit for point processes via topological data analysis.
\newblock \emph{Electronic Journal of Statistics}, 14\penalty0 (1):\penalty0 1024--1074, 2020.
\newblock \doi{10.1214/20-EJS1683}.

\bibitem[Botnan and Hirsch(2022)]{botnan2022}
M.~B. Botnan and C.~Hirsch.
\newblock On the consistency and asymptotic normality of multiparameter persistent {Betti numbers}.
\newblock \emph{Journal of Applied and Computational Topology}, 2022.
\newblock ISSN 2367-1734.
\newblock \doi{10.1007/s41468-022-00110-9}.

\bibitem[Cressie(1993)]{cressie1993}
N.~A.~C. Cressie.
\newblock \emph{Statistics for Spatial Data}.
\newblock Probability \& Mathematical Statistics S. John Wiley \& Sons, 2 edition, 1993.

\bibitem[Diggle(1979)]{diggle1979}
P.~J. Diggle.
\newblock On parameter estimation and goodness-of-fit testing for spatial point patterns.
\newblock \emph{Biometrics}, 35\penalty0 (1):\penalty0 87--101, 1979.
\newblock ISSN 0006-341X.
\newblock \doi{10.2307/2529938}.

\bibitem[Fasy et~al.(2024)Fasy, Kim, Lecci, Maria, Millman, and Rouvreau.]{TDA}
B.~T. Fasy, J.~Kim, F.~Lecci, C.~Maria, D.~L. Millman, and V.~Rouvreau.
\newblock \emph{{TDA}: Statistical Tools for {Topological Data Analysis}}, 2024.
\newblock URL \url{https://CRAN.R-project.org/package=TDA}.
\newblock R package version 1.9.1.

\bibitem[Fend and Redenbach(2025)]{fend2025}
C.~Fend and C.~Redenbach.
\newblock Goodness-of-fit tests for spatial point processes: A review, 2025.
\newblock URL \url{https://arxiv.org/abs/2501.03732}.

\bibitem[Heinrich-Mertsching et~al.(2024)Heinrich-Mertsching, Thorarinsdottir, Guttorp, and Schneider]{heinrichmertsching2024}
C.~Heinrich-Mertsching, T.~L. Thorarinsdottir, P.~Guttorp, and M.~Schneider.
\newblock Validation of point process predictions with proper scoring rules.
\newblock \emph{Scandinavian Journal of Statistics}, 51\penalty0 (4):\penalty0 1533--1566, 2024.
\newblock \doi{10.1111/sjos.12736}.

\bibitem[Krebs and Hirsch(2022)]{krebs2022}
J.~Krebs and C.~Hirsch.
\newblock Functional central limit theorems for persistent {Betti numbers} on cylindrical networks.
\newblock \emph{Scandinavian Journal of Statistics}, 49\penalty0 (1):\penalty0 427–454, 2022.
\newblock ISSN 1467-9469.
\newblock \doi{10.1111/sjos.12524}.

\bibitem[Lavancier et~al.(2015)Lavancier, Møller, and Rubak]{lavancier_determinantal_2015}
F.~Lavancier, J.~Møller, and E.~Rubak.
\newblock Determinantal point process models and statistical inference.
\newblock \emph{Journal of the Royal Statistical Society Series B: Statistical Methodology}, 77\penalty0 (4):\penalty0 853--877, 2015.
\newblock ISSN 1369-7412.
\newblock \doi{10.1111/rssb.12096}.
\newblock URL \url{https://doi.org/10.1111/rssb.12096}.

\bibitem[Loosmore and Ford(2006)]{loosmore2006}
N.~B. Loosmore and E.~D. Ford.
\newblock Statistical inference using the {G} or {K} point pattern spatial statistics.
\newblock \emph{Ecology}, 87\penalty0 (8):\penalty0 1925--1931, 2006.
\newblock \doi{10/bgpbgs}.

\bibitem[M{\o}ller and Waagepetersen(2003)]{moller2003}
J.~M{\o}ller and R.~P. Waagepetersen.
\newblock \emph{Statistical Inference and Simulation for Spatial Point Processes}.
\newblock CRC Press, 2003.
\newblock ISBN 978-0-203-49693-0.

\bibitem[Mrkvi{\v{c}}ka et~al.(2017)Mrkvi{\v{c}}ka, Myllym{\"a}ki, and Hahn]{mrkvicka2017}
T.~Mrkvi{\v{c}}ka, M.~Myllym{\"a}ki, and U.~Hahn.
\newblock Multiple {Monte Carlo} testing, with applications in spatial point processes.
\newblock \emph{Statistics and Computing}, 27\penalty0 (5):\penalty0 1239–1255, 2017.
\newblock ISSN 1573-1375.
\newblock \doi{10.1007/s11222-016-9683-9}.

\bibitem[Mrkvi{\v{c}}ka et~al.(2022)Mrkvi{\v{c}}ka, Myllym{\"a}ki, Kuronen, and Narisetty]{mrkvicka2022}
T.~Mrkvi{\v{c}}ka, M.~Myllym{\"a}ki, M.~Kuronen, and N.~N. Narisetty.
\newblock New methods for multiple testing in permutation inference for the general linear model.
\newblock \emph{Statistics in Medicine}, 41\penalty0 (2):\penalty0 276–297, 2022.
\newblock ISSN 1097-0258.
\newblock \doi{10.1002/sim.9236}.

\bibitem[Myllym{\"a}ki and Mrkvi{\v{c}}ka(2020)]{myllymaki2020}
M.~Myllym{\"a}ki and T.~Mrkvi{\v{c}}ka.
\newblock Comparison of non-parametric global envelopes, 2020.
\newblock URL \url{https://arxiv.org/abs/2008.09650}.

\bibitem[Myllym{\"a}ki and Mrkvi{\v{c}}ka(2024)]{GET}
M.~Myllym{\"a}ki and T.~Mrkvi{\v{c}}ka.
\newblock {GET}: {Global} {Envelopes} in {R}.
\newblock \emph{Journal of Statistical Software}, 111:\penalty0 1--40, 2024.
\newblock ISSN 1548-7660.
\newblock \doi{10.18637/jss.v111.i03}.

\bibitem[Myllym{\"a}ki et~al.(2015)Myllym{\"a}ki, Grabarnik, Seijo, and Stoyan]{myllymaki2015}
M.~Myllym{\"a}ki, P.~Grabarnik, H.~Seijo, and D.~Stoyan.
\newblock Deviation test construction and power comparison for marked spatial point patterns.
\newblock \emph{Spatial Statistics}, 11:\penalty0 19--34, 2015.
\newblock ISSN 2211-6753.
\newblock \doi{https://doi.org/10.1016/j.spasta.2014.11.004}.

\bibitem[Myllym{\"a}ki et~al.(2017)Myllym{\"a}ki, Mrkvi{\v{c}}ka, Grabarnik, Seijo, and Hahn]{myllymaki2017}
M.~Myllym{\"a}ki, T.~Mrkvi{\v{c}}ka, P.~Grabarnik, H.~Seijo, and U.~Hahn.
\newblock Global envelope tests for spatial processes.
\newblock \emph{Journal of the Royal Statistical Society: Series B (Statistical Methodology)}, 79\penalty0 (2):\penalty0 381--404, 2017.
\newblock \doi{10.1111/rssb.12172}.

\bibitem[Ripley(1977)]{ripley1977}
B.~D. Ripley.
\newblock Modelling spatial patterns.
\newblock \emph{Journal of the Royal Statistical Society: Series B (Methodological)}, 39\penalty0 (2):\penalty0 172–192, 1977.
\newblock ISSN 2517-6161.
\newblock \doi{10.1111/j.2517-6161.1977.tb01615.x}.

\bibitem[Robins and Turner(2016)]{robins2016}
V.~Robins and K.~Turner.
\newblock Principal component analysis of persistent homology rank functions with case studies of spatial point patterns, sphere packing and colloids.
\newblock \emph{Physica D: Nonlinear Phenomena}, 334:\penalty0 99–117, 2016.
\newblock ISSN 0167-2789.
\newblock \doi{10.1016/j.physd.2016.03.007}.

\end{thebibliography}

\clearpage

\begin{appendices}

\section{Continuous rank ordering for the 1-dimensional Betti curve}\label{sec:cont-details}

As mentioned in Section~\ref{sec:result-indivi}, the continuous rank ordering (cont) performs worse than the other orderings when the selected functional summary statistic is the $1$-dimensional Betti curve $\beta_1$. In this section, we discuss a small example which illustrates the problem.

The continuous rank measure was introduced in \citet{mrkvicka2022}. The idea behind this ordering is to identify the extremeness of a curve via local extremeness over a small subset of the domain. Below, we provide the brief construction of the continuous rank.

Let $D_i = [D_i^{r_1}, \dots, D_i^{r_n}]$ denote the pointwise values of the discretized $\beta_1$ curve of the $i$th point pattern at $n$ distinct evaluation points $r_1 < r_2 < \dots < r_n$. In total, we have $m\!+\!1$ vectors corresponding to $i=0,\dots,m$ where $i=0$ denotes the observed point pattern. We want to order these vectors using the continuous rank ordering.

For this ordering, first denote by $D_{(0)}^r \leq \dots \leq D_{(m)}^r$ the increasingly ordered $m\!+\!1$ values at evaluation point $r$.

If there are no ties involving the $i$th ordered value, then its raw continuous pointwise rank is defined to be
\begin{equation}
    c_{(i)}^r = \begin{cases}
    \exp\!\left(-\frac{D_{(1)}^r - D_{(0)}^r}{D_{(m)}^r - D_{(1)}^r}\right)\cdot\1{D_{(1)}^r < D_{(m)}^r} & \text{if } i=0,\\
    i + \frac{D_{(i)}^r - D_{(i-1)}^r}{D_{(i+1)}^r - D_{(i-1)}^r} & \text{if } 0 < i < m,\\
    m\!+\!1 - \exp\!\left(-\frac{D_{(m)}^r - D_{(m-1)}^r}{D_{(m-1)}^r - D_{(0)}^r}\right)\cdot\1{D_{(0)}^r < D_{(m-1)}^r} & \text{if } i=m.
    \end{cases} \label{eq:point-wise-rank-1}
\end{equation}

If there are ties for the $i$th ordered value, i.e. there are some $k\neq l$ with $0 \leq k \leq i \leq l \leq m$ such that $k = \min\{0 \leq k' \leq i \mid D_{(k')}^r = D_{(i)}^r\}$ and $l = \max\{i \leq l' \leq m \mid D_{(l')}^r= D_{(i)}^r\}$, then the raw continuous rank is defined as \begin{equation}
    c_{(i)}^r = \frac{k+l+1}{2}. \label{eq:point-wise-rank-2}
\end{equation}

In our goodness-of-fit test, the pointwise continuous rank of curve $i$ at evaluation point $r$ is given as 
\begin{equation}
C_i^{r} =  \min(c_i^r, m\!+\!1 - c_i^r) \label{eq:point-wise-rank-3}
\end{equation}
to take both large and small deviations into account.

In the next step, the pointwise continuous ranks are summarized in the continuous rank for the entire vector by taking the minimum of the pointwise ranks, i.e.,
\begin{equation}
    C_i = \min_{r \in \{r_1, \dots, r_n\}} C_i^{r} \quad \text{for } \quad i=0,\dots,m.
\end{equation}
Small values of $C_i$ correspond to the most extreme curves, while the most central curves get the largest continuous ranks.

\begin{figure}[th]
    \centering
    \hfill
    \subfloat[point pattern $\x_0$]{\includegraphics[height=5cm]{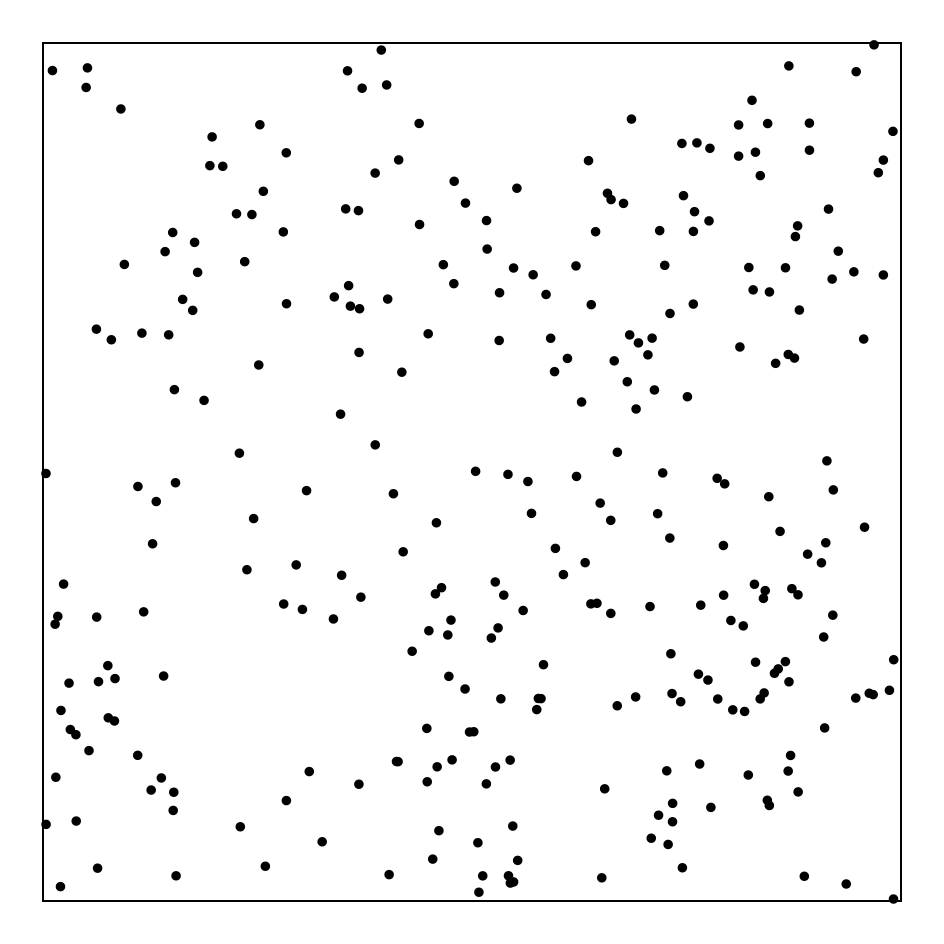}\label{fig:ex-matclu-pattern}}
    \hfill
    \subfloat[$20$ empirical $\beta_1$ curves used in the test]{\includegraphics[height=5cm]{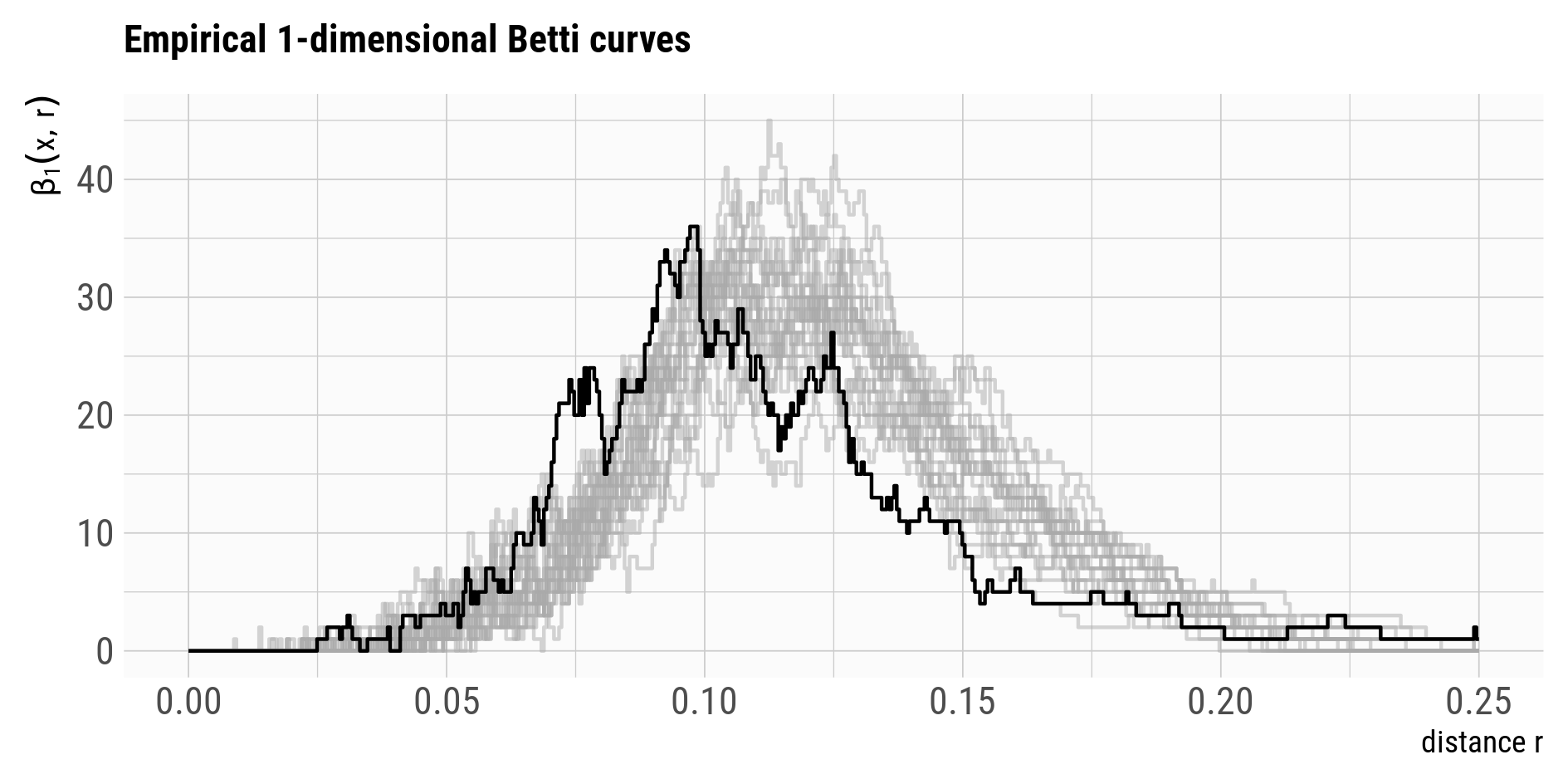}\label{fig:ex-matclu-curves}}

    \subfloat[graphical representation of the erl ordering]{\includegraphics[width=0.5\linewidth]{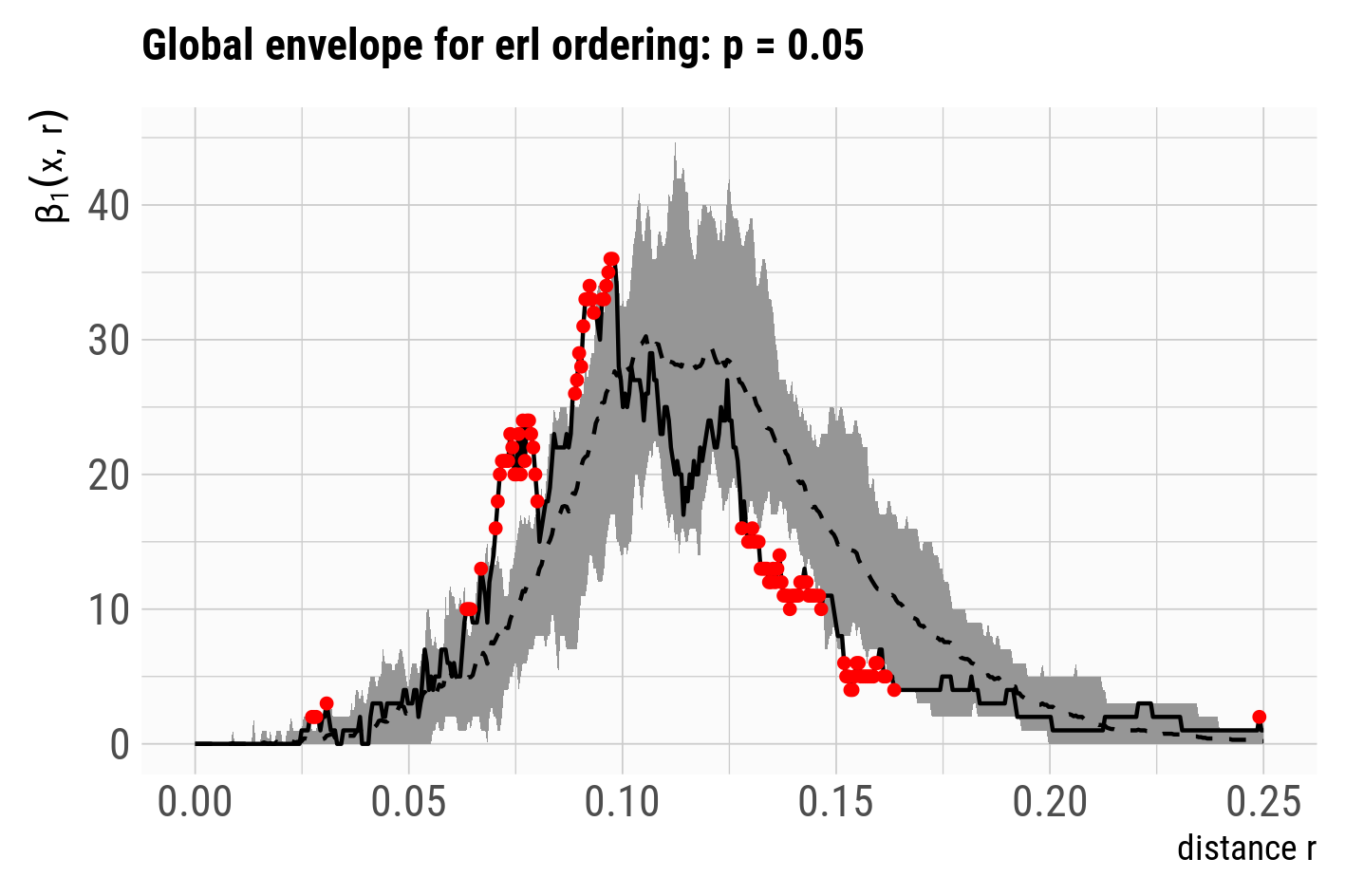}\label{fig:ex-matclu-erl}}
    \subfloat[graphical representation of the cont ordering]{\includegraphics[width=0.5\linewidth]{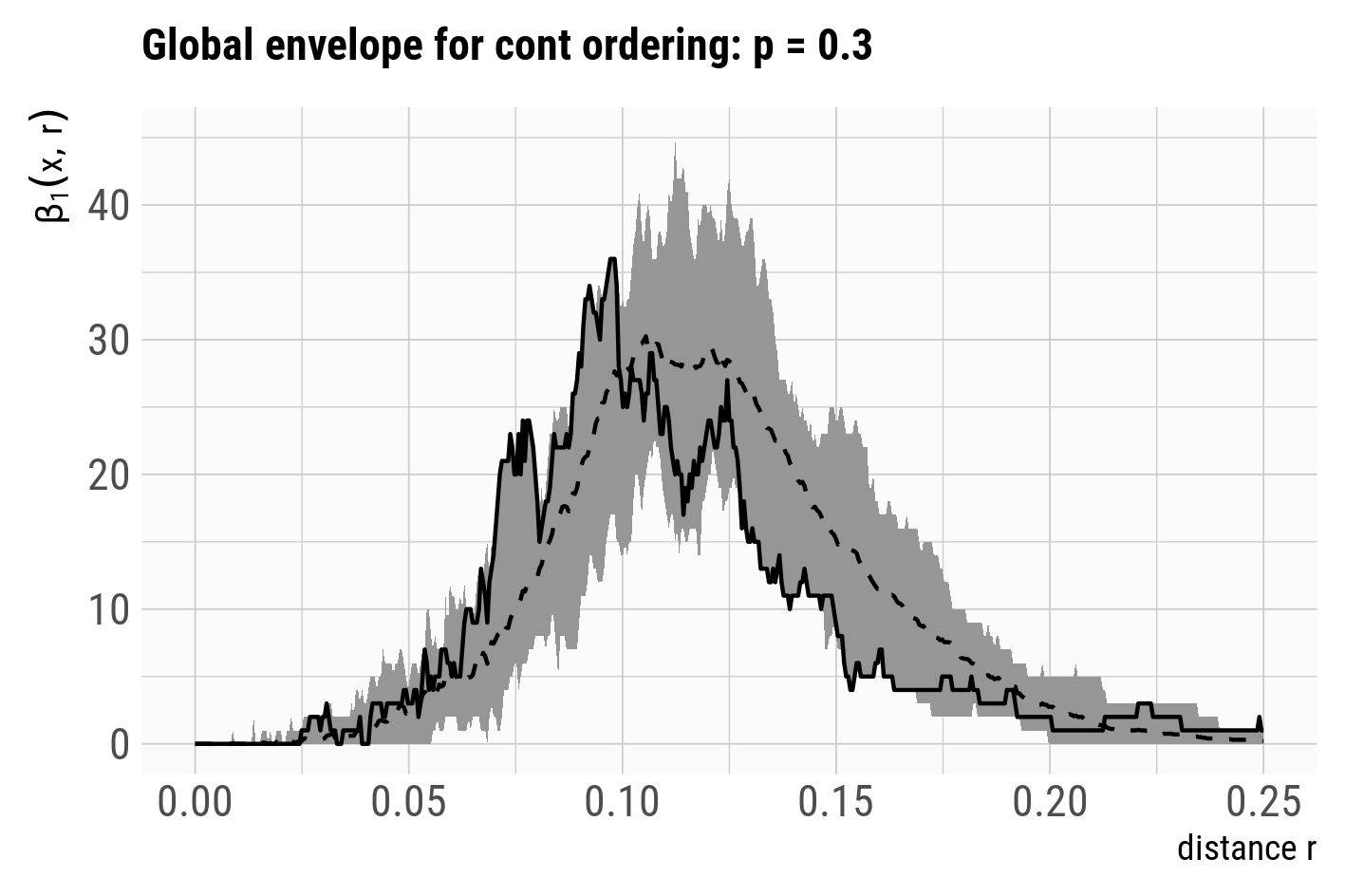}\label{fig:ex-matclu-cont}}

    \caption{Visualization of the global envelope tests for the null hypothesis of CSR using the $1$-dimensional Betti curve as functional summary statistic and either the erl or the cont ordering. The observed pattern shown in (a) is a realization of the MatClu model on the observation window $W_6$. We used $m=19$ simulations of CSR. All $20$ curves are shown in (b) where the black solid line is the curve obtained from $\x_0$. The global envelopes corresponding to the test using FUN and the erl ordering ($p\text{-value} = 0.05$, rejection) and the one with the cont ordering ($p\text{-value} = 0.3$, acceptance) are shown in (c) and (d), respectively. The dashed line corresponds to the pointwise mean of the Betti curves. The red dots indicate which distances led to the rejection of the null hypothesis.}
    \label{fig:comp-envelopes}
\end{figure}

In our example, we consider one of the realizations of the MatClu model on the observation window $W_6$. In our study, we were not able to reject the null hypothesis of CSR for this pattern when using the cont ordering, regardless of the number of simulations $m$. The corresponding pattern $\x_0$ is shown in Figure~\ref{fig:ex-matclu-pattern}. For demonstration purposes, we used only $m=19$ simulations of the CSR null model in the Monte Carlo tests illustrated in the following. Since $m=19$, we reject the null hypothesis for $\x_0$ if and only if $\beta_1(\x_0, \cdot)$ is the single most extreme curve. 

The $\beta_1$ curves for the $20$ point patterns used in the tests are shown in Figure~\ref{fig:ex-matclu-curves}. The solid black line corresponds to the observed curve $\beta_1(\x_0, \cdot)$.  

\begin{figure}[th]
    \centering
    \includegraphics[width=\linewidth]{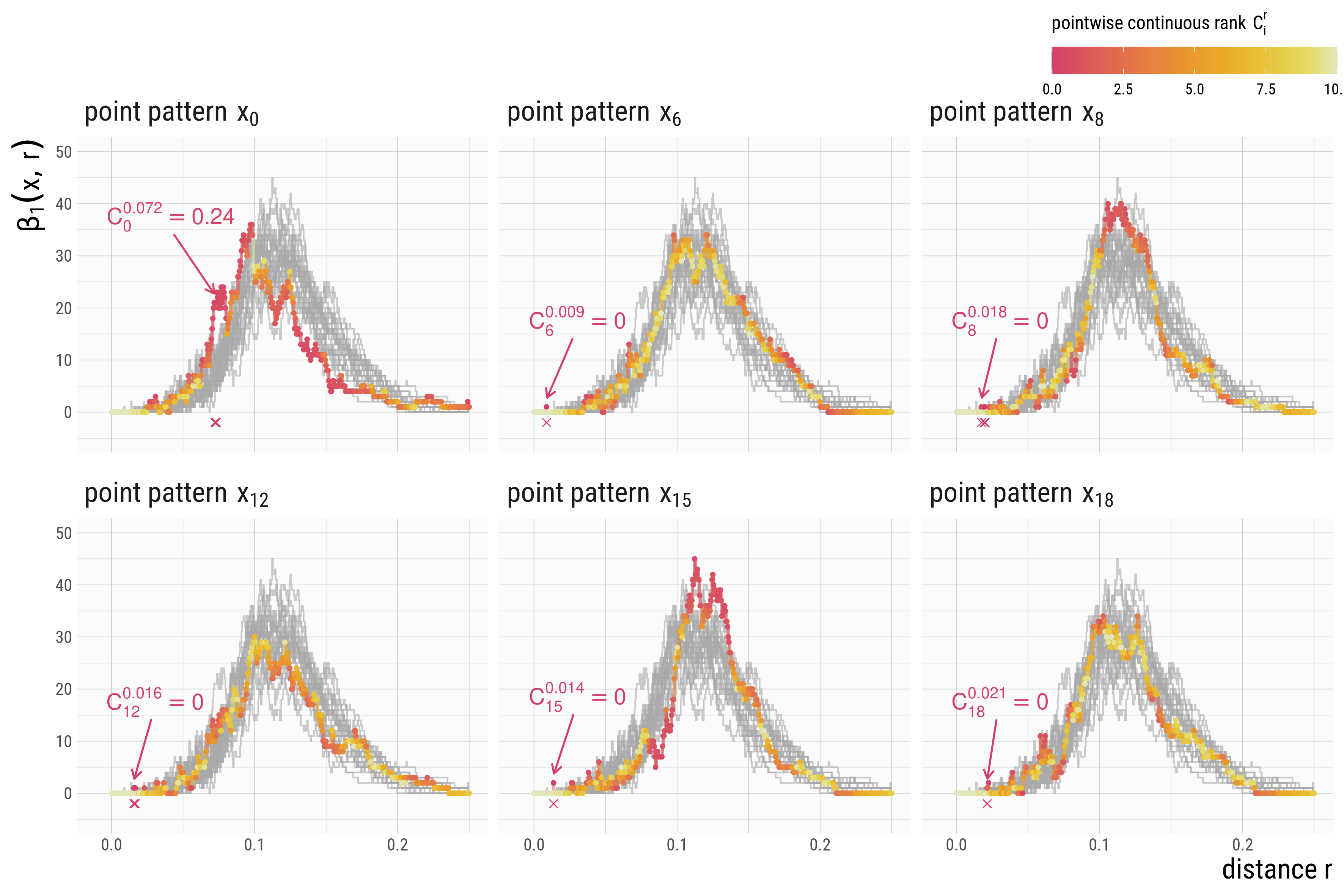}
    \caption{Pointwise continuous ranks $C_i^{r}$ of th $1$-dimensional Betti curves $\beta_1(\x_i, \cdot)$. Pattern $\x_0$ is the observed point pattern, patterns $\x_6, \x_{8}, \x_{12}, \x_{15}$ and $\x_{18}$ have a smaller continuous rank than $\x_0$ and are hence more extreme. Small pointwise continuous ranks shown in red are hereby extreme while the yellow coloring with $C_i^r = 10$ corresponds to the most central pointwise values. The red crosses correspond to all evaluation points where $C_i^r \leq \min_r C_0^r = 0.24$ (the continuous rank measure of the observed point pattern).}
    \label{fig:expl-cont}
\end{figure}

The graphical representations in terms of the global envelopes of the tests with test statistic FUN and the erl and the cont ordering are shown in Figure~\ref{fig:ex-matclu-erl} and Figure~\ref{fig:ex-matclu-cont}, respectively.
In case of the extreme rank length ordering, we reject CSR, as the solid line leaves the gray envelope at at least one evaluation point ($p_{\operatorname{MC}}=0.05$). For the continuous ranks, we accept the null hypothesis ($p_{\operatorname{MC}}=0.3$). For this ordering, $5$ point patterns (patterns $\x_6, \x_{8}, \x_{12}, \x_{15}$ and $\x_{18}$) of the $19$ simulated patterns are deemed more extreme than $x_0$. Figure~\ref{fig:expl-cont} shows for the six relevant point patterns all pointwise continuous ranks of the $1$-dimensional Betti curves, as well as the evaluation point with minimal pointwise continuous rank. 

Let $i$ be the index of one the five simulated curves that are extreme under the cont ordering. There exists at least one evaluation point $r$ for each curve where the pointwise continuous rank $C_i^r$ for $\beta_1(\x_i, r)$ is equal to $0$. 
This results for each of the five curves in the overall value of the continuous rank of $C_i = \min_r C_i^r = 0$ which is the smallest possible continuous rank that can be achieved.

A continuous rank of $0$ is also the reason that the observed curve coincides with the boundary of the global envelope for a large part of the domain. 

For the global envelope as graphical representation of the test with significance level $\alpha$, we first need to compute a threshold value $\nu_\alpha$. This threshold is the largest value of the individual continuous ranks such that the number of the continuous ranks strictly smaller than $\nu_\alpha$ is at most $\alpha(m+1)$ (see Eq.~(51) in \citet{fend2025} for details).
In our example with $m=19$, significance level $\alpha=0.05$ and the existence of one continuous rank of $0$ we obtain the threshold $\nu_\alpha=0$. The global envelope shown in Figure~\Cref{fig:ex-matclu-cont} is finally formed by taking the pointwise minimal and maximal value of all curves where $C_i \geq \nu_\alpha = 0$. In our case, this results in taking the extreme values of all $20$ $1$-dimensional Betti curves, the $19$ simulated ones and the observed one.

We will now look in more detail at the cases in which $C_i^r = 0$ occurs. From \eqref{eq:point-wise-rank-1}, \eqref{eq:point-wise-rank-2} and \eqref{eq:point-wise-rank-3} it becomes clear that $C_i^r = 0$ can only be achieved if there are no ties involved, as otherwise the raw continuous rank $c_i^r$ will never be $0$ or $m+1$ due to the averaging. Additionally, due to the definition in \eqref{eq:point-wise-rank-1} we see that the value of the corresponding Betti curve at point $r$, i.e. $D_i^r$, needs to be either the largest or the smallest value, thus we are in the first or the last case. Since the exponential term is strictly positive, the indicator needs to be zero to get a raw continuous rank of either $0$ or $m+1$. We conclude in both cases that all the remaining $m$ ordered values $D_{(1)}^r, \dots, D_{(m)}^r$ or $D_{(0)}^r, \dots, D_{(m-1)}^r$ need to coincide.

Thus, we only have two appearing values $d_1 \neq d_2$, with $m$ curves attaining the one value $d_1$ and one curve (in our case the $i$th one) attaining the other value $d_2$. This yields the pointwise and overall continuous rank for the $i$th curve of $0$. Note that in this scenario, neither the absolute nor the relative difference between the two values is taken into account. In our example, we have $d_1 = 0$ and either $d_2=1$ (for the patterns $\x_{6}, \x_{8}, \x_{10}, \x_{18}$) or $d_2 = 2$ (pattern $\x_{15}$). Thus, the entire curves are considered more extreme because there exists some evaluation point where one or two $1$-dimensional topological holes formed. These holes do not persist for a long time and would be considered topological noise in topological data analysis. 

\begin{figure}[th]
    \centering
    \includegraphics[width=\linewidth]{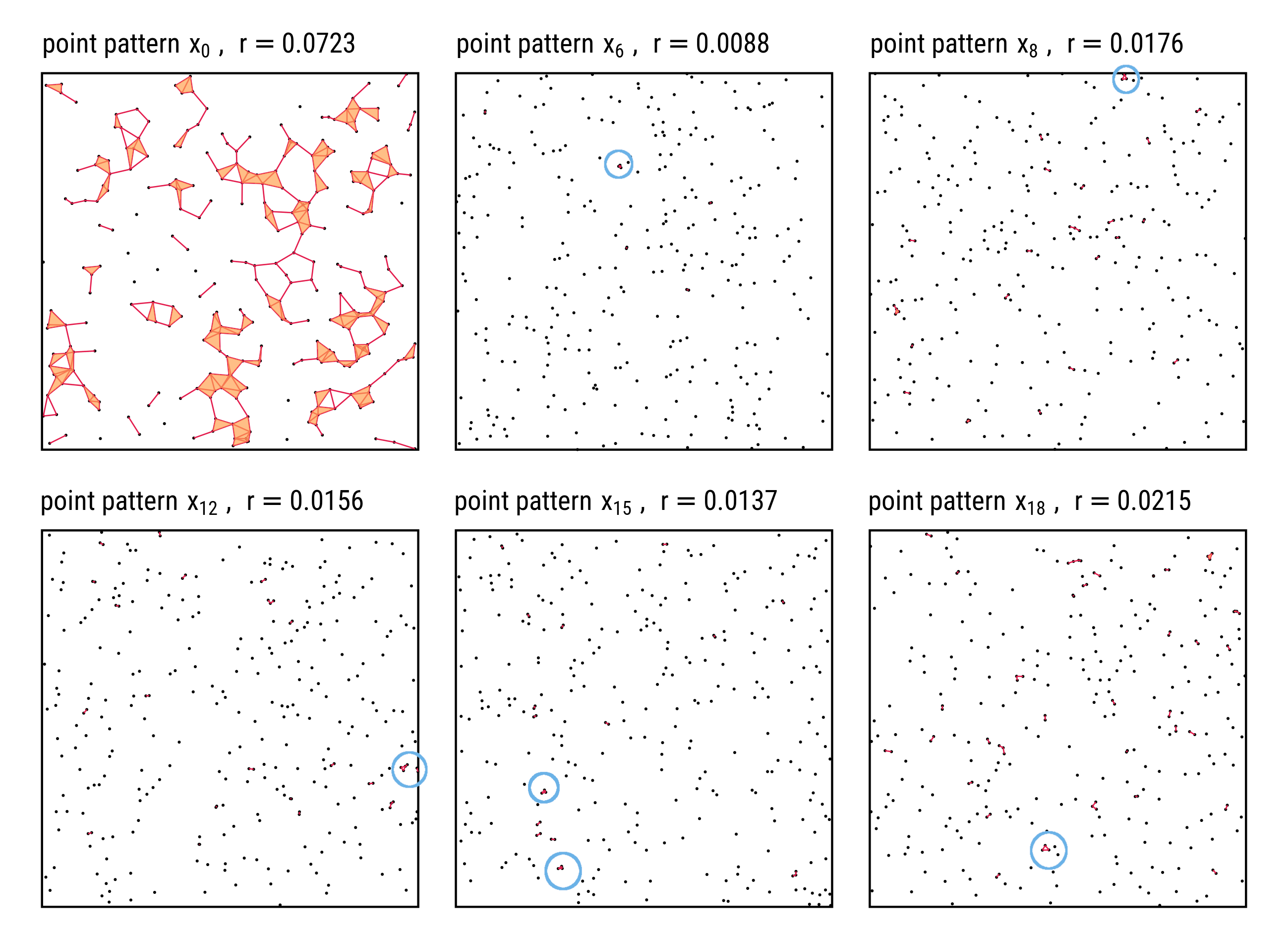}
    \caption{Alpha-complexes built on top of the point patterns at distance $r$. The distance for each pattern matches the distance where the minimal continuous rank of the $1$-dimensional Betti curve is attained. The individual $1$-dimensional topological features (circular holes) that exist in the complex are highlighted in blue in case they are hard to identify by eye.}
    \label{fig:sim-patterns-complex}
\end{figure}

The corresponding alpha-complexes at the respective \emph{extreme} evaluation points (as annotated in Figure~\ref{fig:expl-cont}) highlighting the small $1$-dimensional topological features are shown in Figure~\ref{fig:sim-patterns-complex}.

This example illustrates why we need a higher number of simulations for the cont ordering -- in contrast to the erl and area ordering --  when the functional summary statistics has evaluation points where it is very likely that almost all curves attain the same value. 

\end{appendices}

\end{document}